\documentclass[%
 twocolumn,
 amsmath,amssymb,
 aps, physrev,
]{revtex4-2}

\usepackage{graphicx}
\usepackage{dcolumn}
\usepackage{bm}
\usepackage{physics}
\usepackage{amsmath}
\usepackage{comment}
\usepackage{bibunits}

\begin{document}

\title{\textbf{A Bifurcation Theory for the Equilibria of Modern Hopfield Networks}}%

\author{Vincenzo Maria Schimmenti$^{\dagger,1}$}
\author{Matteo Ciarchi$^{\dagger,1,2,3}$}%
\email{Contact author: mc2663@cam.ac.uk}
\affiliation{
$^{1}$Max-Planck-Institute for the Physics of Complex Systems, Noethnitzer Str. 38, 01187 Dresden, Germany \\
$^{2}$Ludwig-Maximilians-Universit\"at M\"unchen, Arnold-Sommerfeld-Center for Theoretical Physics, Theresienstr. 37, 80333 M\"unchen, Germany
}
\affiliation{$^{3}$Gurdon Institute, University of Cambridge, Cambridge CB2 1QN, UK}

\date{\today}

\begin{abstract}

Modern Hopfield networks provide a unifying framework for associative memory, transformer attention, diffusion-based generative models, and biological attractor dynamics, linking these systems through a common energy-based dynamics in which states are updated toward weighted combinations of stored patterns. Across these settings, network dynamics is determined by the organization of the energy landscape and the bifurcations of its fixed points. Despite their central role, a general theory of these bifurcations has remained unavailable beyond specific architectures and idealized pattern ensembles. Here we derive stability and bifurcation criteria for the fixed points of general convex-dual Modern Hopfield networks for a general statistics of the stored patterns. Applying this framework to random, block-correlated, and infinitely hierarchical pattern ensembles, we show how memory correlations systematically organize the emergence of hierarchical attractors through successive bifurcations. We further demonstrate that these predictions quantitatively describe retrieval bifurcations in MHNs storing patterns sampled from MNIST, and recapitulate the hierarchical organization of hematopoietic cell identities. Our results establish a general bifurcation theory for Modern Hopfield networks and identify the organization of fixed points as a unifying principle underlying their computational and biological behavior.

\end{abstract}

\maketitle

\begin{bibunit}[apsrev4-2]

Modern Hopfield networks (MHNs) are associative memory architectures that exploit continuous states and nonlinear softmax interactions to achieve memory capacities far exceeding those of the classical Hopfield model \cite{hopfield1982neural,krotov2016dense,demircigil2017model,ramsauer2020hopfield}. Stored patterns correspond to minima of an energy landscape, and retrieval proceeds through relaxation toward the appropriate attractor \cite{krotov2020large}. Beyond associative memory, MHNs have emerged as a unifying framework across machine learning and biology: their dynamics are mathematically equivalent to transformer attention \cite{vaswani2017attention,ramsauer2020hopfield}, closely related to the denoising dynamics of diffusion models \cite{pham2025memorization}, and arise naturally from the convex Lagrangian formulation of general associative memories, which unifies classical, dense, and hierarchical associative memories within a common energy-based framework \cite{krotov2020large}. Their high storage capacity and generalization-like properties have also been used to classify immune cell types \cite{widrich2020modern}. More recently, gene regulatory circuits governing cell identity have been modeled based on MHNs, where differentiated cell types correspond to attractors and multipotent progenitors emerge as bifurcation-induced mixture states \cite{OmerKarin1, OmerKarin2}.

Across these diverse settings, the computational behavior of an MHN is determined by the organization of its fixed points. As the nonlinearity controlling the sharpness of the softmax is increased, the energy landscape undergoes a cascade of bifurcations that change the attractors. These transitions govern which memories are retrievable, when spurious or mixed states appear, and how hierarchical representations emerge. In random Hopfield models, they underlie the classical retrieval-to-spin-glass transition \cite{amit1985storing, amit1987statistical}; in modern generative models, they govern the emergence and stability of memorized, spurious, and compositional attractors, thereby shaping the transition between memorization and generalization \cite{pham2025memorization, kalaj2025random, ambrogioni2024search, hoover2023memory}. Finally, in developmental systems, they reproduce the successive lineage decisions that generate cellular hierarchies \cite{OmerKarin1}. A general understanding of MHNs therefore requires a theory of how fixed points are created, destabilized, and organized by bifurcations.

Despite their central role, bifurcations in MHNs remain understood only in restricted settings. Existing analyses rely on replica or interpolation techniques tailored to specific energy functions and largely to pre-defined pattern ensembles \cite{achilli2025capacity, lucibello2024exponential, beise2026unstable}. Beyond analyses of correlated and hierarchically organized attractors in specific models \cite{gutfreund1988hierarchical,tivno2007equilibria,OmerKarin1,OmerKarin2}, no general theory exists for the hierarchy of bifurcations and saddle points generated by arbitrary correlated memory ensembles. A general stability and bifurcation theory, applicable to arbitrary pattern statistics and capable of addressing real hierarchical datasets, is still lacking.

In this work, we derive general stability and bifurcation criteria for the fixed points of MHNs valid for general statistics of the stored patterns. By employing a convex-dual formulation of the MHN energy, this framework reveals how the correlation structure of memories determines the hierarchy of attractors and the sequence of bifurcations that organize the energy landscape. We apply it to i.i.d. memories, block-correlated ensembles, and a continuum limit with infinitely many hierarchical levels, which exposes the asymptotic structure of the bifurcation cascade itself. Finally, we demonstrate that the same theoretical framework quantitatively predicts retrieval bifurcations on MHN with patterns sampled from MNIST, and recovers the hematopoietic lineage tree from single-cell data. Together, these results establish bifurcation theory as a general organizing principle for the dynamics and representation learned by modern Hopfield networks.

\section{Convex-dual formulation of modern Hopfield networks}


Here, we consider the pattern retrieval dynamics given by a modern Hopfield network (MHN) in an $N$-dimensional feature space of vectors $\mathbf{x} = (x_1,x_2...,x_N)^T$. We introduce a matrix $\boldsymbol{\Xi}$, which is a $K\times N$ matrix whose components $\Xi_{\mu i}$ indicate the $i$-th feature value of pattern $\mu$. In the following, we will denote each pattern as a vector with $\boldsymbol{\xi}_\mu$. The dynamics of features follow the equation:
\begin{equation}
    \dot{\boldsymbol{x}} = \boldsymbol{\Xi}^\intercal \operatorname{softmax}(\beta \boldsymbol{\Xi} \boldsymbol{x}) - \boldsymbol{x} \,
    \label{eq:dynamics}
\end{equation}
where $\operatorname{softmax}(\boldsymbol{\theta})$, for $\boldsymbol{\theta} \in \mathbb{R}^K$, is a K-dimensional vector of components $\operatorname{softmax}(\boldsymbol{\theta})_\mu=\frac{e^{\theta_\mu}}{\sum_\nu e^{\theta_\nu}}$. While the formulation of \eqref{eq:dynamics} naturally uses the dot product to assess the similarity between features and patterns, it is useful to introduce the "Euclidean" version of an MHN via the following dynamics:
\begin{equation}
    \dot{\boldsymbol{x}} = \boldsymbol{\Xi}^\intercal \operatorname{softmax}\left(-\frac{\beta}{2} ||\boldsymbol{x} - \boldsymbol{\Xi}||^2\right) - \boldsymbol{x} \,
    \label{eq:dynamics_eucl}
\end{equation}
We introduce this modified version of the MHN because, for large $\beta$ and arbitrary pattern norms, the dynamics tends to bias toward patterns with large norms. For equally normalized patterns, this bias has no effect. In the general case, it is always possible, formally, to obtain \eqref{eq:dynamics_eucl} from \eqref{eq:dynamics}
by introducing an "energetic" bias:
\begin{equation}
    \dot{\boldsymbol{x}} = \boldsymbol{\Xi}^\intercal \operatorname{softmax}(\beta \boldsymbol{\Xi} \boldsymbol{x} + \beta \boldsymbol{a}) - \boldsymbol{x} \,
    \label{eq:dynamics2}
\end{equation}
and choosing $a_\mu = -\frac{1}{2} ||\boldsymbol{\xi}_\mu||^2$. We refer to the bias $\boldsymbol{a}$ as "energetic" since it vanishes for $\beta=0$. An "entropic" bias, on the other hand, would be $ \ beta$-independent. We leave more details on this distinction in the Supplementary Materials, keeping the form of dynamics as in \eqref{eq:dynamics2} as our reference. 
It is well known that the dynamics in \eqref{eq:dynamics} and \eqref{eq:dynamics2} can be obtained by a gradient descent dynamics for the potential $V(\boldsymbol{x}) = \frac{1}{2} ||\boldsymbol{x}||^2 - \frac{1}{\beta} \log \left(\sum_{\mu=1}^K e^{\beta (\boldsymbol{\xi}_\mu^\intercal \boldsymbol{x} + a_\mu) }\right)$ \cite{ramsauer2020hopfield}. The fixed points of Eq.~\eqref{eq:dynamics2} constitute the patterns retrieved under the dynamics.

By introducing the convex dual of the log-sum-exp function, it is possible to recast the fixed-point equation of  Eq.~\eqref{eq:dynamics2} as an optimization problem for a simplex variable $\boldsymbol{w}$ (see Supplementary Materials). Stable fixed points are then given by the maxima of a convex-dual functional:
\begin{equation}
\max_{\boldsymbol{w} \in \Delta_K} \Phi_\beta(\boldsymbol{w}), \quad 
\Phi_\beta(\boldsymbol{w}) = \frac{1}{2}\boldsymbol{w}^\top \boldsymbol{G} \boldsymbol{w} + \boldsymbol{w}^\intercal \boldsymbol{a} + \frac{1}{\beta}H(\boldsymbol{w}),
\label{eq:freeenergyw}
\end{equation}
where $H(\boldsymbol{w}) = -\sum_{\mu=1}^K w_\mu \log w_\mu$, and $\boldsymbol{w} \in \Delta_K = \{w_\mu \ge 0, \sum_{\mu=1}^K w_\mu = 1\}$,. The quadratic term encodes pattern correlations via the patterns' Gram matrix $\boldsymbol{G}=\boldsymbol{\Xi}\boldsymbol{\Xi}^\intercal$, while the entropy term, $H(\boldsymbol{w})$, favors uniform states. Interestingly, this free energy has the structure of a multi-component Flory--Huggins free-energy density \cite{flory1942thermodynamics,huggins1941solutions,bartolucci2026metastabilityripeningmulticomponentliquid} (see also Supplementary Materials). An equivalent fixed-point dynamics in dual space is:
\begin{equation}
    \dot{\boldsymbol{w}} = \operatorname{softmax}\left(\beta (\boldsymbol{G} \boldsymbol{w} + \boldsymbol{a}) \right)  - \boldsymbol{w}
\end{equation}
This shows that the fixed points of the retrieval dynamics in the pattern simplex space are uniquely determined by the Gram matrix of the patterns, the value of $\beta$, and the bias $\boldsymbol{a}$. The maxima of the convex-dual are thus determined by a competition between an "energetic" term $e(\boldsymbol{w})=\frac{1}{2}\boldsymbol{w}^\top\boldsymbol{G}\boldsymbol{w}+\boldsymbol{w}^\top\boldsymbol{a}$ and an entropic term $-\frac{1}{\beta}H(\boldsymbol{w})$.

\paragraph*{Stability of fixed points}
We now study the stability behavior of the fixed points of the MHN in the convex-dual formulation. For $\beta=0$ the fixed point of both the standard and dual dynamics is given the average of patterns, i.e. $\boldsymbol{x}^* = \frac{1}{K}\boldsymbol{\Xi}^\intercal \boldsymbol{1}_K = \frac{1}{K}\sum_{\mu=1}^K \boldsymbol{\xi}_\mu$ and $\boldsymbol{w}^* = \frac{1}{K} \boldsymbol{1}_K$, where $\boldsymbol{1}_K$ denotes the $K$-dimensional vector of all ones. As $\beta$ changes, the unique fixed point changes continuously until the first bifurcation at a critical $\beta$.
The stability of all fixed points can be studied via the Jacobian of the dual dynamics:
\begin{equation}
    \boldsymbol{J}_w = \beta \boldsymbol{F}(\boldsymbol{w}) \boldsymbol{G} - \boldsymbol{I}
    \label{eq:Jacobian}
\end{equation}
with $\boldsymbol{F}(\boldsymbol{w}) = \operatorname{diag}(\boldsymbol{w}) - \boldsymbol{w} \boldsymbol{w}^\intercal$ the Fisher matrix related to the simplex element $\boldsymbol{w}$. Stability of a fixed point $\boldsymbol{w}^*$ thus requires:
\begin{equation}
    \lambda_{\rm \max}(\boldsymbol{J}_w) < 0 \implies \beta < \frac{1}{\lambda_{\rm max}(\boldsymbol{F}(\boldsymbol{w}^*) \boldsymbol{G})}
\end{equation}
For numerical purposes, the most tractable form is:
\begin{equation}
    \beta < \frac{1}{\lambda_{\rm max}(\boldsymbol{F}(\boldsymbol{w}^*)^{1/2} \boldsymbol{G} \boldsymbol{F}(\boldsymbol{w}^*)^{1/2})}
\end{equation}
For the uniform fixed point introduced above, this corresponds to:
\begin{equation}
    \beta < \beta_c = \frac{K}{\lambda_{\rm max}\left( \left(\boldsymbol{I}_K - \frac{1}{K} \boldsymbol{1} \boldsymbol{1}^\intercal\right) \boldsymbol{G}\right)}.
\end{equation}
Hence, general bifurcations happen at those values of $\beta$ at which the largest eigenvalue of the Gram matrix, projected to the orthogonal space of $\boldsymbol{w}^*=\frac{1}{K} \boldsymbol{1}$, makes the Jacobian cross zero when the uniform state is protected as a stable fixed point at low $\beta$. As we will see, the critical $\beta_c$ obtained in this way still gives a good estimate for the retrieval transition, even when, for $\beta\neq0$, the uniform state is not a fixed point anymore.
The region of stability of the uniform fixed point can be studied by expanding the fixed point condition $\boldsymbol{w} = \operatorname{softmax}(\beta( \boldsymbol{G} \boldsymbol{w} + \boldsymbol{a}))$ around a stable fixed point $\boldsymbol{w}_0 \equiv \boldsymbol{w}^*(\beta_0)$ at $\beta_0$. This gives:
\begin{equation}
    (\boldsymbol{I} - \beta_0 \boldsymbol{F}(\boldsymbol{w}_0) \boldsymbol{G}) \boldsymbol{w}_0' = \boldsymbol{F}(\boldsymbol{w}_0) \left(\boldsymbol{G} \boldsymbol{w}_0 + \boldsymbol{a}\right)
\end{equation}
where $\left. \boldsymbol{w}_0'=\frac{d\boldsymbol{w}^*(\beta)}{d\beta}\right|_{\beta=\beta_0}$. Applying this to $\beta_0=0$ we can immediately see that if $\boldsymbol{a} = - \frac{1}{K} \boldsymbol{G} \boldsymbol{1}_K$ then $\boldsymbol{w}_0' = \boldsymbol{0}$ and the uniform fixed point is stable for small $\beta-\beta_0$. This also occurs, in the case without bias $\boldsymbol{a} = \boldsymbol{0}$, if the patterns are centered: in this case $\sum_{\mu} G_{\mu \nu} = \sum_\nu G_{\mu \nu} = 0$.

As the uniform state becomes unstable, the first bifurcation occurs along the eigenvector whose eigenvalue crosses zero. Its direction will determine the features on which the new fixed point will concentrate. Further bifurcations are determined analogously, where again the directions of the unstable eigenvectors of the $\boldsymbol{G}$ matrix restricted to the sub-simplex via $\boldsymbol{F}(\boldsymbol{w}^*)$ (Eq.~\eqref{eq:Jacobian}) determine further localizations in pattern space. This can be illustrated by considering a state localized on a subset of patterns $B\subset\{1,...,K\}$ in the simplex space. Assume that $\sum_{i\notin B}w_i=\delta\ll1$ and write $w_B=(1-\delta)\hat{w}$, with $\sum_{i\in B}\hat{w}_i=1$. Then for $G$ bounded the stability operator obeys
\begin{equation}
    F(w)G = \begin{pmatrix} F(\hat{w})G_{BB} & 0 \\
    0 & 0
\end{pmatrix} + \mathcal{O}(\delta).
\end{equation}
The stability condition thus selects the subspace concentrated on the localized patterns. In particular, if the selected block $B$ has a logit gap:
\begin{equation}
    \max_{j\notin B}(Gw+a)_j\leq\max_{i\in B}(Gw+a)_i-\epsilon_{\text{logit}},
\end{equation}
then $\sum_{j\notin B}w_j\lesssim(K-|B|)e^{-\beta\epsilon_{\text{logit}}}$, $|B|$ cardinality of $B$. This shows that localization requires only linear separation in logit space. Once localized, the same arguments for the bifurcation of the selected block as discussed above apply.

The type of the local bifurcation at a critical $\beta_c$ can be probed by considering an expansion of the convex-dual potential around $\beta_c$ in orders of the coefficient of the most unstable eigenvector $\boldsymbol{\hat{t}} \in T = \{ \delta\boldsymbol{w}:\boldsymbol{1}^\top \delta\boldsymbol{w}=0\}$. Writing $\boldsymbol{w} = \boldsymbol{w}_0+\epsilon \boldsymbol{\hat{t}}$, for a simple instability of a smooth parent branch persisting through $\beta_c$, one obtains an effective potential as a function of $\epsilon$ (see Supplementary Materials for complete derivation and expression of the coefficients $A_i$):
\begin{equation}
    \Phi^{\text{eff}}_{\beta}(\epsilon) = \Phi_{\beta}^0+ A_1 \epsilon + \frac{A_2}{2} \left(\frac{1}{\beta} - \frac{1}{\beta_c}\right) \epsilon^2 + \frac{A_3}{3} \epsilon^3 + \dots  .
    \label{eq:effectiveLandauz}
\end{equation}
If $A_1 = 0$, the non-zero fixed point of~\eqref{eq:effectiveLandauz} is given by $\epsilon^* = \frac{A_2}{A_3} \frac{\beta-\beta_c}{\beta_c \beta}$, and the daughter branch follows the unstable direction linearly in $\beta-\beta_c$ through a transcritical bifurcation; the parent branch is maintained.

If $A_1\neq 0$, the uniform fixed point shifts; the original fixed point is only metastable, and the fixed point position changes with $\beta$. If symmetry imposes $A_1=0$ and $A_3=0$, then the fourth-order term needs to be retained, and $\epsilon^*\propto\pm\sqrt{\beta-\beta_c}$ provided $A_4\neq 0$: two daughter branches emerge from the initial fixed points through a pitchfork bifurcation. For a generic branch that does not persist through the instability, a linear unfolding term is allowed, and the corresponding local bifurcation is a fold \cite{kuznetsov2004elements} (see the Supplementary Materials). Importantly, the coefficients $A_i$ depend on high-order statistics of the data, like its dominant PCA eigenvector or its kurtosis (see Supplementary Materials). The nature of the first transition thus depends on the geometrical distribution of the data.
Once the first bifurcation happens, we can repeat the argument for the new fixed point localized on a subset of the patterns. Under the assumption stated above of logit linear separation, we expect this to reproduce the observed transitions qualitatively.

This analysis thus shows that the local bifurcations of an MHN are organized in "symmetry-breaking" transitions that localize the fixed points on patterns with increasing correlations. This is similar to what is observed in diffusion models, where a symmetry-breaking event defines the emergence of generalization \cite{raya2024spontaneous}.
The Landau expansion above characterizes the local loss of stability of a given fixed-point branch and therefore describes bifurcations to states continuously connected to it. It does not, however, preclude discontinuous transitions between distinct fixed-point branches. In a hierarchical memory ensemble, several localization levels may coexist as local extrema of the dual functional, and the globally selected state is determined by their relative values of $\Phi_\beta$. Consequently, an intermediate level can be skipped if it never becomes the global maximum: the system then jumps directly between two separated localization scales before the parent branch necessarily loses local stability. This distinction between local instability and global branch crossing is the hierarchical analogue of the spinodal/coexistence distinction in first-order phase transitions and underlies the Maxwell construction obtained below in the continuum limit.

\section{Random and block-structured patterns}

\begin{figure}
    \centering
    \includegraphics[width=\columnwidth]{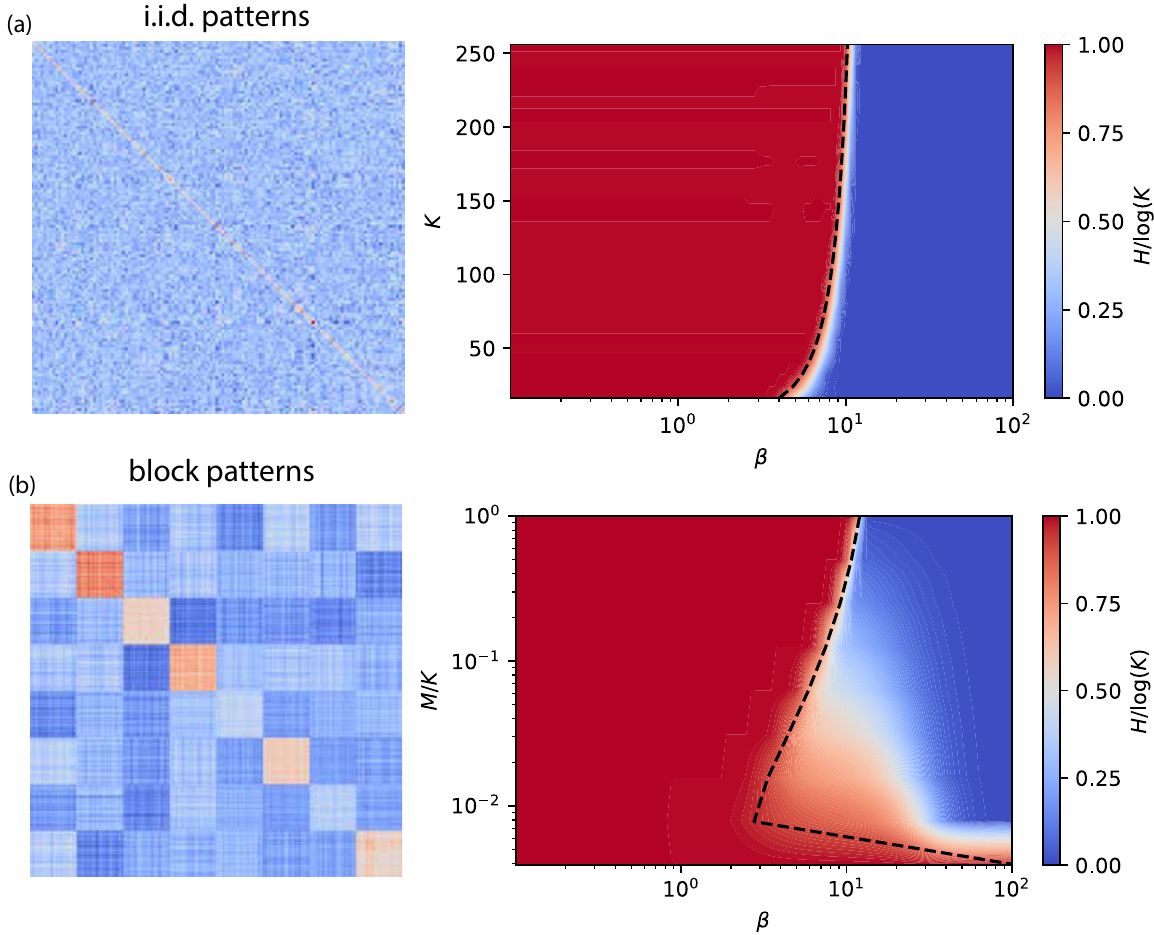}
    \caption{(a) Gram matrix (left) and entropy as a function of $\beta$ and number of patterns $K$. In the Gram matrix, red indicates high correlation, while blue indicates low correlation. The color scale in the heatmap on the right indicates the pattern entropy normalized by $\log K$. The dashed line in the heatmap indicates the transition line as predicted by the crossing given the random-matrix theory (RMT) eigenvalues. (b) Example of a Gram matrix for eight blocks (left) and entropy as a function of $\beta$ and the ratio of the number of blocks to the number of patterns $M/K$. The color scale indicates the pattern entropy normalized by $\log K$. We use inter-blocks correlation $\rho_0 = 0.1$ and intra-block $\rho_1 = 0.9$. The dashed line in the heatmap indicates the transition line as predicted by stability analysis on the uniform state.}
    \label{fig:Fig2IIDBlock}
\end{figure}

We now apply the results of the previous sections to the analysis of the bifurcation structure in some exemplary models.
At first, we study the emergence of a hierarchy of fixed points for two examples: random patterns and a block-structured Gram matrix. To characterize the changes in fixed point structure, we compute the dependence of the pattern entropy $H(\boldsymbol{w})= -\sum_{\mu=1}^K w_\mu\log w_\mu $ on $\beta$. This quantity constitutes a good probe of the localization of the fixed points on the pattern features: at low $\beta$, the uniform pattern corresponds to the maximum entropy. As $\beta$ increases, the entropy decreases until it reaches the minimum value of $0$ once fixed points localize on single patterns. At the intermediate bifurcations, we expect the entropy to change non-analytically in $\beta$.

We first consider $K$ i.i.d. Gaussian-distributed patterns in $N$ dimensions. We draw the patterns independently as $\boldsymbol{\xi}_\mu$ as $\mathcal{N}(\boldsymbol{0}_N, \frac{\boldsymbol{I}_N}{N})$. The Gram matrix is a Wishart Matrix with $K$ rows and aspect ratio $\alpha=K/N$ (example in Fig.~\ref{fig:Fig2IIDBlock}(a), left). To locate the first instability (of the uniform state), we study the largest eigenvalue of $\boldsymbol{F}(\boldsymbol{w}_0) \boldsymbol{G}$. As the orthogonal projection to the uniform point affects only the smallest eigenvalues, we can drop the effect of $\boldsymbol{F}(\boldsymbol{w}_0)$ and study the largest eigenvalue of $\boldsymbol{G}$. When $K$ and $N$ are large, we expect corrections to the eigenvalue distribution to be small, and we can approximate the largest eigenvalue with the right support limit of the Marchenko--Pastur law \cite{marchenko1967distribution}. This gives (see Supplementary Materials):
\begin{equation}
    \beta_{c, \rm iid} \approx \frac{K}{(1+\sqrt{K/N})^2}
\end{equation}
As we discussed, the composition of the fixed point at low $\beta$ strongly depends on the presence of a regularizing bias (Euclidean), the presence of an additional bias, and the centering of the patterns. To appreciate this difference, we report the entropy as a function of $\beta$ for the normal MHN (Fig.\ref{fig:Fig2IIDBlock}) (the Euclidean case is shown in the Supplementary Materials). Fig.~\ref{fig:Fig2IIDBlock}(a), right, shows that the stability criterion obtained from the eigenvalue statistics reproduces well the transition line as a function of $\beta$ and $K/N$. Furthermore, we observe no hierarchy of bifurcations, but a single transition from a unique fixed point to multiple fixed points (as quantified by the pattern entropy). This is intuitively given by the unstructured distribution of patterns in the i.i.d. case.  

We then consider a system of $K=\alpha M$ patterns with unit modulus arranged in $M$ blocks of size $\alpha$. We set the intra-block correlations to $\rho_1$, and the inter-block correlation to $\rho_0$. The symmetry of the problem allows for an analytical treatment of the bifurcations. By minimizing the free energy in $\mathbf{w}$ space in Eq.~\eqref{eq:freeenergyw}, we can identify the critical $\beta$ values at which the fixed points change support (see Supplementary Materials). The first transition happens from a uniform $\mathbf{w}$ among the patterns to fixed points concentrated inside the single blocks. The second transition involves a further localization to fixed points concentrated in single patterns. Notably, for specific values of $M/K$ and $\alpha$, the second temperature is lower than the first one, meaning that there is a unique bifurcation from a uniform state to states localized on single patterns. To verify the validity of these arguments in a real-case scenario, we analyze the transitions in stable fixed points for random block-Gram matrices (an example of a Gram matrix for 8 blocks is in Fig.~\ref{fig:Fig2IIDBlock}(b), left). We obtained the stable fixed points numerically for different values of $\beta$ and $M/K$ (Fig.~\ref{fig:Fig2IIDBlock}(b)), where we show the entropy of the patterns for different $\beta$ values and different numbers of blocks. We observe two transitions for large block sizes (small $M/K)$, while at small block sizes the transition becomes unique. Intuitively, this can be explained by the behavior in the limit of a single block ($G_{\mu \nu} = c~\forall~\mu,\nu$): in this limit, the matrix is a matrix with equal entries, and all patterns are equal. The transition in this case is thus unique. The criterion on the first bifurcation (instability of the uniform fixed point) reproduces well the first transition (dashed line in Fig.~\ref{fig:Fig2IIDBlock}(b), right).

These two regimes demonstrate that hierarchical retrieval is governed by spectral separation in $G$: clustered spectra produce staged symmetry breaking and metastability, whereas featureless spectra lead to abrupt, effectively one-step selection.

To gain insight into the general case, we study the continuum limit of a hierarchical Gram matrix (see Supplementary Materials). We introduce the layer variable $s\in[0,1]$ obtained from the limit of $l/L,~l\in \{0,..., L\}$ as $L\rightarrow\infty$ (the ratio of the layer index $l$ to the total number of layers $L$, where $l=0$ corresponds to the layer associated with the full matrix). Every layer is defined by the size of the blocks in the layer, $N(s)$, and the overlap of the patterns inside the same block in the layer, $q(s)$. In this limit, the selected layer over which sites localize at a value of $\beta$ is given by the maximization of the layer free energy:
\begin{equation}
    \mathcal{F}(s) = \beta^{-1}\log{N(s)}+\frac{1}{2}\left[ q(s)+\int_s^1\frac{N(u)}{N(s)}q'(u)du \right].
    \label{eq:FreeEnergylayercont}
\end{equation}
The presence of a hierarchy of bifurcations depends on the interplay between the entropy penalty due to the localization on a block at a definite level $s$ and the gain from the overlap of the patterns in the same block. Differentiating Eq.~\eqref{eq:FreeEnergylayercont} with respect to $s$ produces:
\begin{equation}
    \mathcal{F}(s)'=\frac{N'(s)}{N(s)}\left[\frac{1}{\beta}-\frac{I(s)}{2} \right],
    \label{eq:minimacontinuumlayer}
\end{equation}
where $I(s) = \int_s^1 \frac{N(u)}{N(s)}q(u)' du$. Under the assumption that $N(s)'< 0$ (decreasing block size in the hierarchy) and $q(s)'>0$ (increasing correlation for higher layers), if $I(s)' < 0~\forall~s $ then Eq.~\eqref{eq:minimacontinuumlayer} has a unique solution $s^*$ for fixed $\beta$, and $\frac{ds^*}{d\beta}>0$, meaning that as $\beta$ increases higher layers are selected continuously. By introducing the level energy $E(s) = \frac{1}{2}(q(s)+I(s))$, the free energy in Eq.~\eqref{eq:FreeEnergylayercont} can be written as $\mathcal{F}(s) = \beta^{-1}\log N(s)+E(s)$. In this formulation, a change in $\beta$ is equivalent to adding a tilt to the curve $(\log N(s), e(s))$: since $\frac{d E(s)}{d \log N(s)} = -1/2 I(s)$, $I(s)' < 0~\forall~s $ is equivalent to requiring that $E(\log N(s))$ is a concave function of the entropy. If this is not satisfied, the selected layer changes discontinuously, analogously to what is obtained from a Maxwell construction. This gives a quantitative criterion for the determination of the hierarchical bifurcations in large systems with hierarchical correlations.

\section{Analysis of MNIST dataset}

\begin{figure*}
    \centering
    \includegraphics[width=\linewidth]{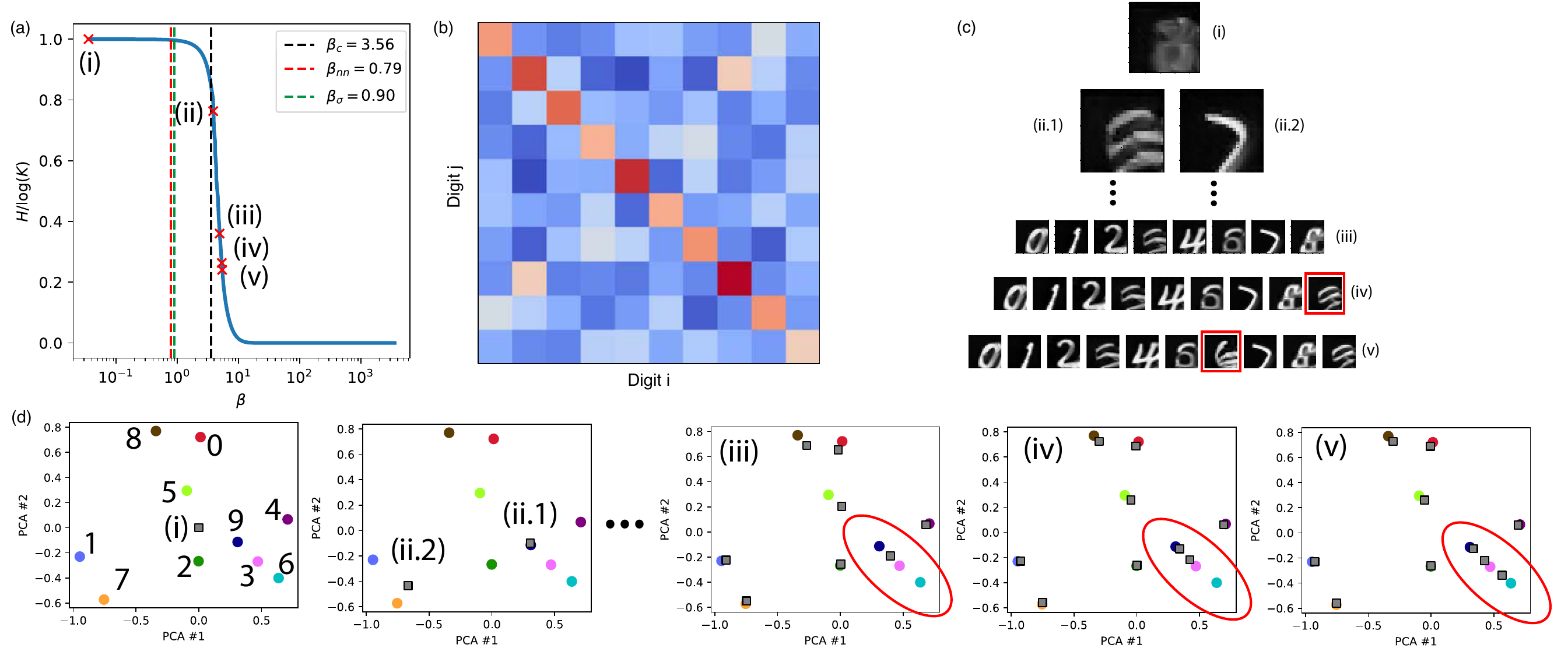}
    \caption{Analysis of the MNIST dataset, with 1 sample per digit (label). (a) Pattern entropy dependence on $\beta$. We indicated the estimated transitions based on the distribution variance compared to the pattern variance and nearest-neighbor distance with dashed lines (green and red, respectively). The black dashed line indicates the estimated $\beta_c$ at which the uniform fixed point becomes unstable. Roman numerals indicate the $\beta$ values at which we plotted the fixed points in (b). (b) Gram matrix of the fixed points. Red indicates high correlation, while blue indicates low correlation. (c) Images of the samples of fixed points of the MHN dynamics at four values of $\beta$, in increasing order of $\beta$. (d) First two PCA coordinates of the retrieved fixed points (squares) and the stored patterns (circles). We indicated the corresponding digit above the pattern points. The red ellipses indicate the group of patterns closest to the selected fixed points in red squares in (c).}
    \label{fig:Fig3MNIST}
\end{figure*}

To test the validity of our analysis on real data, we now consider MNIST as an example. Classically, when looking at the retrieval capabilities of the associated network, research was devoted to using the pixel-space data directly as memories. While this is more direct, it suffers greatly from the spatial nature of this type of data, due to the competition between different features and their spatial distribution. In order to disambiguate the two, we first map the images into a compressed feature space, employing an Image Autoencoder. Autoencoders are machine learning models that learn a factorized identity map between a space and itself. Concretely, one trains an encoder $\boldsymbol{z} = \boldsymbol{f}_\theta(\boldsymbol{x})$ and a decoder $\boldsymbol{x}' = \boldsymbol{g}_{\theta'}(\boldsymbol{z})$ trying to minimize the reconstruction error $\frac{1}{2} ||\boldsymbol{x'}-\boldsymbol{x}||^2$. In particular, we employ a pre-trained Variational Autoencoder \cite{kingma2014autoencoding} from Stability AI (available on HuggingFace Hub). The idea is that we can leverage the autoencoder for compressing the images to a latent space, which is more amenable to analysis with an MHN. An image space analysis would be possible, but it would require endowing the MHN with a spatial kernel or embedding the image pixels into "visual tokens" and adding positional embeddings, which goes beyond the scope of our analysis. It is worth stressing that such an approach is not solely relevant for retrieval, but also for generative models, in the language of latent-space diffusion \cite{rombach2022latent}.

For MNIST, since the Autoencoder expects an RGB image as input, we replicate the grayscale channel across the three RGB channels present in the dataset images, mimicking the three colors. As usual, MNIST contains 10 labels \cite{lecun1998gradient}. As patterns, we randomly choose 1, 2, and 5 patterns per label to construct the dataset. This keeps the dataset balanced and avoids "entropic" effects due to a relative abundance of a class with respect to another \cite{nicoletti2026interplay}.

For the analysis of the fixed points, we leverage the dual formulation of the free energy in Eq.~\eqref{eq:freeenergyw} in the Euclidean case. We set as patterns the digits obtained from the autoencoder in latent space. We start by initializing the weights on the vertices of the pattern simplex, corresponding to a $\beta \to \infty$ state: for pattern $\mu^*$, we set $w_\mu = \delta_{\mu,\mu^*}$. We then perform an instantaneous quenching to a selected $\beta$. Different values of $\beta$ probe the bifurcation structure at different hierarchical levels. For high $\beta$ values, the dynamics' initial point is essentially already at a fixed point, while for values in between bifurcations, many initial conditions converge to the same fixed point. To better individuate the fixed point, we map back weights $\boldsymbol{w}^*$ to real-space patterns $\boldsymbol{x}^*$ and spatially cluster them (we moreover remove duplicate fixed points arising from different initial conditions). We analyzed the fixed-point entropy at varying $\beta$ (Fig.~\ref{fig:Fig3MNIST}(a)). We observe a gradual decrease in entropy, corresponding to an increase in the localization of the fixed points on the patterns. We also estimated the critical value $
\beta_c$ at which the initial unique fixed point loses stability: this matches quantitatively with the observed initial dip in entropy. We also estimate the values of $
\beta$ at which the variance of the equilibrium distribution matches the variance of all the data $\beta_{\sigma}$ and the square of the average nearest-neighbor distance ($\beta_{\text{nn}}$). For the MNIST dataset, these values precede the instability of the uniform fixed point. The hierarchy of bifurcation, as discussed before, is determined by the Gram matrix (\ref{fig:Fig3MNIST}(b)). We expect bifurcation to concentrate progressively fixed points on correlated patterns. This is already visible by looking at the fixed-point digits as selected $\beta$ values (Fig.~\ref{fig:Fig3MNIST}(c)). While the first fixed point is a mixture of different digits, increasing values of $\beta$ generate more defined mixtures, which correspond to averages of correlated patterns. Note that the transition is initially dominated by digits with higher diagonal elements of the Gram matrix, and patterns with lower diagonal entries tend to be resolved later (digits 3, 5, 6, and 9, for example). The concentration of fixed points on clouds of correlated patterns is also observable in PCA space (Fig.~\ref{fig:Fig3MNIST}(d)). There, we observe a progressive localization of fixed points on neighboring stored patterns. This shows that the hierarchy of bifurcations concentrates the fixed points around correlated clouds of stored patterns (Fig.~\ref{fig:Fig3MNIST}(d)). Note, however, that the bifurcation sequence seems to concentrate the fixed points on single patterns immediately, and the entropy decreases quite smoothly as a function of $\beta$. This is due to the simple structure of the Gram matrix, which resembles more an i.i.d. matrix.

\section{Bifurcation structure of cell lineages in hematopoiesis}

As a further practical example of the emergence of hierarchical bifurcations, we consider the bifurcation of lineages during hematopoiesis (emergence of blood cell types during development) \cite{OmerKarin1, OmerKarin2}. In this context, the emergence of cell types during development has been successfully explained through a model whose dynamics follows the dynamics of MHNs \cite{OmerKarin1}. In this model, groups of co-activated enhancers constitute cell types, which define the patterns of the corresponding MHN. The emergence of new cell types happens at bifurcation points of the MHN.

\begin{figure}
    \centering
    \includegraphics[width=\columnwidth]{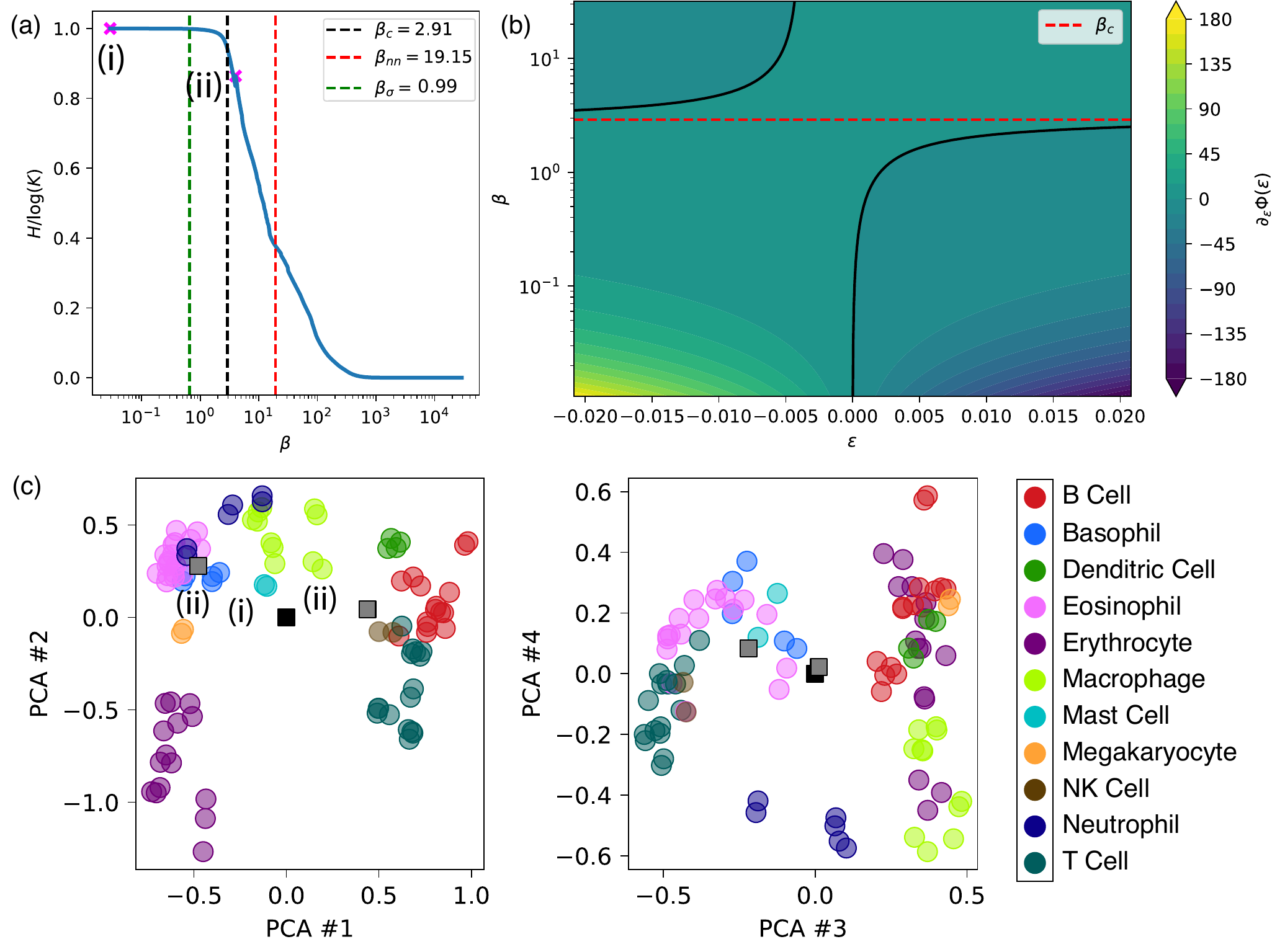}
    \caption{Analysis of the hematopoiesis dataset. (a) Pattern entropy dependence on $\beta$. The position of the dashed lines is calculated as in Fig.~\ref{fig:Fig3MNIST}. (b) Gradient of the dual potential (color scale) in the plane $(\epsilon, \beta)$. In black are the zero-gradient level lines. The red dashed line indicates the critical $\beta$ at which the first fixed point bifurcates. (c) First four PCA coordinates of the patterns (circles) and fixed points (squares) at the three values of $\beta$ indicated in (a).  Circles and fixed points are colored based on the cell lineage they belong to (names on the right). The black square denoted by (i) corresponds to the uniform fixed point, which bifurcates into the two gray squares. The bifurcations are indicated in the entropy plot in (a) via the crosses. As discussed, the first bifurcation occurs mostly along the first principal component, while the contribution from the others remains more modest.  }
    \label{fig:Fig4HEMA}
\end{figure}

We initialized the Euclidean MHN by inserting, as stored patterns, the expression patterns of the lineages found in the Haemopedia dataset \cite{choi2019haemopedia}, with genes selected as in \cite{OmerKarin1}.
According to the discussion above, we expect a cascade of bifurcations at varying $\beta$ with a fixed-point structure determined by the cell type correlation matrix. This cascade should proceed through fixed points whose pattern weights $\mathbf{w}$ concentrate on subsets of cell types with higher correlation as $\beta$ increases, until reaching single-pattern fixed points. This is confirmed by the dependence of the pattern entropy on $\beta$ (Fig.\ref{fig:Fig4HEMA}(a), Supplementary Materials). We observe different decay behaviors of the entropy as $\beta$ increases, indicating the presence of different bifurcations. The entropy curve shows more regimes than the MNIST dataset one, hinting at a more complex hierarchy of bifurcations. The estimated $\beta_\sigma$ and $\beta_{\text{nn}}$ illustrate the presence of at least two transitions. 

To study the stability of the uniform fixed point, we consider perturbations of the effective potential in the convex-dual formulation expanded around it (Fig.~\ref{fig:Fig4HEMA}(b)). Starting from the uniform weight over patterns, we perturb the state by a quantity $\epsilon$ along the most unstable eigenvector of the Jacobian matrix, which is dual (in the SVD sense) to the first principal component. By looking at the lines of zero gradient, we can determine when the uniform fixed point loses stability. Around $\beta \approx \beta_c$, the two lines disconnect, emphasizing the splitting of the uniform fixed point.

As we will show below, the bifurcations organize in a hierarchical cascade that localizes the fixed points on patterns with higher correlation. This is further confirmed by looking at the fixed point position in PCA space (Fig.~\ref{fig:Fig4HEMA}(c)). As always, at low $\beta$ values, the system has a unique fixed point at the average of all patterns. An increase in $\beta$ causes fixed points to relocate inside the point clouds of groups of highly correlated patterns. The structured nature of the Gram matrix makes the bifurcation hierarchy richer than in the MNIST dataset.

\begin{figure*}
    \includegraphics[width=\linewidth]{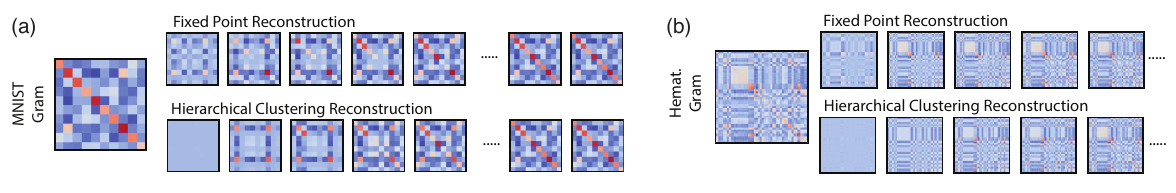} %
    \caption{(a) Gram matrix of the MNIST digits. Upper blocks: from left to right, at each bifurcation, we run a linear regression and find $\boldsymbol{A}$ as the minimizer of $||\boldsymbol{\Xi} - \boldsymbol{A} \boldsymbol{X}^*||^2$, where $\boldsymbol{X}^*$ are the fixed point as some $\beta$ (increasing from left to right). We then report the Gram matrix of the reconstructed patterns, $\boldsymbol{\Xi}^* {\boldsymbol{\Xi}^*}^\intercal $. Lower blocks: same pattern reconstruction procedure, where we employ as basis $\boldsymbol{X}^*$ the clusters obtained by direct hierarchical clustering at increasing tree depth. 
    (b) Same as in (a), but for the Hematopoiesis dataset for selected values of $\beta$.}
    \label{fig:Fig5Recon}
\end{figure*}

\section{Reconstruction of the hierarchy of bifurcations}

In the example of the block Gram matrices in Section II, we showed that bifurcations localize the fixed-point components on blocks of highly correlated patterns. This hints at the possibility that a correlation-sensitive clustering of the stored patterns, like hierarchical clustering, can actually approximate well the series of transitions observed in MHNs. To test this idea, we employ a reconstruction method that aims at obtaining the pattern vectors from the observed fixed points via a linear projection. To do so, we introduce the projection matrix $\boldsymbol{A}^*$ that minimizes the norm $||\boldsymbol{\Xi} - \boldsymbol{A}\boldsymbol{X}^*||$ over $K\times K^*$ matrices $\boldsymbol{A}$, $K^*$ number of fixed points. Intuitively, the matrix $\boldsymbol{A}^*$ gives the best linear reconstruction of the stored patterns from the observed fixed points of the MHN at a fixed $\beta$. We then obtain the reconstructed Gram matrix $\Xi^* = \boldsymbol{A}^* \boldsymbol{X}^*$. The same procedure can be carried out by considering the basis of hierarchical clusters $\boldsymbol{X}_h^*$ for the different hierarchical tree levels. This is a $K_h\times N$ matrix, where $K_h$ is the number of clusters at the selected level. The comparison of the resulting Gram matrices indicates how similar the vector spaces spanned by the fixed point at a $\beta$ value and the hierarchical clusters at a hierarchy level are. This is easily illustrated by considering $P$ linearly independent fixed points: in this case $\boldsymbol{X}\boldsymbol{X}^T$ is invertible, and the optimal $\boldsymbol{A}$ is given by $\boldsymbol{A}^* = \boldsymbol{\Xi}\boldsymbol{X^{*\top}}(\boldsymbol{X^{*}}\boldsymbol{X^{*\top}})^{-1}$. Consequently, the effective Gram matrix is $\boldsymbol{\Xi}^* = \boldsymbol{\Xi}\text{Proj}_{\boldsymbol{X}^*}\boldsymbol{\Xi}^\top$, where $\text{Proj}_{\boldsymbol{X}^*}$ is the projection on the subspace spanned by $\boldsymbol{X^{*}}$. Similar Gram matrices, therefore, indicate similar projection subspaces between the fixed points and the clusters of a hierarchy level.

We carried out this procedure for the MNIST (Fig.~\ref{fig:Fig5Recon}(a)) and the hematopoiesis (Fig.~\ref{fig:Fig5Recon}(b)) datasets. The corresponding reconstructed Gram matrices show that the hierarchy of fixed points at varying $\beta$ is well reproduced by the clusters obtained via hierarchical clustering.
The transitions observed for MNIST mostly reproduce the order of appearance of the fixed points by the value of the Gram diagonal entry. This is due, as discussed above, to the low structure of the MNIST dataset Gram matrix.
For the hematopoiesis data, the reconstruction shows an emergence of correlated groups of fixed points, where these localized to groups of patterns with increasing correlation. This indicates that the MHN is able to reproduce the differentiation hierarchy observed for these cell types \cite{OmerKarin2}. Overall, this shows that the bifurcation transitions of MHNs organize in a hierarchy determined by the correlation structure encoded in the Gram matrix.

\section{Discussion and Outlook}

In this work, we have identified general principles governing the organization of bifurcations in Modern Hopfield networks. By exploiting the convex-dual representation of the dynamics, we showed that the stability of a fixed point is controlled by the spectrum of the pattern Gram matrix projected onto the local tangent space of the pattern simplex. This provides a direct connection between the statistical structure of the stored memories and the organization of the corresponding energy landscape: as the inverse temperature $\beta$ is increased, unstable spectral directions determine how fixed points successively localize onto subsets of increasingly correlated patterns. Bifurcations therefore provide a natural mechanism for converting the correlations among stored patterns into a hierarchy of attractors.

We applied the same framework across settings of increasing structural complexity. Hierarchical and ultrametric organization in associative memories has a long history, including models in which correlated pattern ensembles are prescribed so that retrieval proceeds through successive levels of a hierarchy \cite{gutfreund1988hierarchical,rammal1986ultrametricity}. Here, instead, the hierarchy of attractors is read directly from the correlation geometry of a fixed memory ensemble through the bifurcations of the MHN. For i.i.d.\ random patterns, the absence of pronounced correlation structure leads to an essentially direct transition from a delocalized state to pattern-specific retrieval. Introducing block correlations instead produces intermediate attractors supported on groups of memories with increasing correlation, with successive bifurcations resolving progressively finer levels of the imposed hierarchy. The continuum hierarchical construction extends this picture to arbitrarily many levels and makes explicit the competition between the entropic cost of localization and the energetic gain associated with correlations. The analysis of these models shows that spectral separation generates separated scales of symmetry breaking, while a featureless spectrum favors an abrupt collapse toward individual memories.

Importantly, the same organizing principle persists beyond these idealized examples. In MNIST, the bifurcation structure reflects the organization of real data in latent feature space and provides quantitative information about retrieval transitions and their dependence on dataset size. In the hematopoietic dataset, the sequence of fixed-point splittings follows biologically meaningful groups of correlated cell identities, showing how an attractor hierarchy can recover the structure of a developmental lineage \cite{OmerKarin1,OmerKarin2}. Taken together, these examples suggest that bifurcation analysis provides a common language for describing representation and retrieval in settings ranging from synthetic associative memories to machine-learning datasets and biological dynamical systems.

The connection between MHNs and diffusion models \cite{ambrogioni2024search,pham2025memorization,hoover2023memory} raises the possibility that the onset of memorization can be predicted directly from the spectral and bifurcation properties of the dataset. Memorization of individual training examples has been demonstrated directly in diffusion models \cite{carlini2023extracting}, while recent work has related the tendency to memorize to data statistics and to the low-noise part of the denoising process \cite{nicoletti2026interplay,shah2025generation}. In particular, dataset size, dimensionality, correlation structure, and noise level all modify the spectrum of the Gram matrix and therefore the locations and separation of the bifurcation scales. Combined with the correspondence between MHN retrieval and diffusion denoising, this suggests a route toward determining when a generative dynamics ceases to represent collective features of the data distribution and begins to resolve individual samples. Such an approach could turn the generalization--memorization transition into a quantitatively characterizable bifurcation phenomenon rather than a property diagnosed only after training or sampling.

More broadly, the framework developed here suggests that the spectrum and hierarchy of correlations in a dataset encode a corresponding hierarchy of dynamical representations. Establishing how this hierarchy is modified by learned feature maps, finite-sample effects, noise, and the time-dependent dynamics of diffusion models will be important steps toward a general theory connecting associative memory, generative modeling, and biological attractor dynamics. From this viewpoint, bifurcations can be understood as a mechanism by which the statistical organization of data is translated into the organization of an energy landscape. 

At the technical level, the next step is to generalize the present results to an annealing dynamics in $\beta$, for which the quenched-like fixed points studied here describe the accessible basins of attraction, and finally to finite-temperature sampling of the MHN energy landscape, as recently partially explored in \cite{alswaidan2026stochasticattentionlangevindynamics} as "stochastic attention". Such a study should also be implemented by an analysis of the basin of attraction of MHNs, as these will determine the retrieval dynamics starting from an initial condition, important for context involving prompting, for example.
Moreover, as a future direction, one could explore further the connection with Flory--Huggins theory \cite{flory1942thermodynamics,huggins1941solutions} and its multi-component extensions, promoting the MHN to a spatially extended model \cite{bartolucci2026metastabilityripeningmulticomponentliquid}.

\section{Acknowledgments}
We thank Luca Ambrogioni, Enrico Ventura, Beatrice Achilli, Matteo Negri, Flavio Nicoletti and Omer Karin for fruitful discussions. We are particularly grateful to Stefano Mannelli and Flavio Nicoletti for organizing the conference "Mathematical Foundations of AI" where this work at an early stage was presented. V.M.S. acknowledges the support of the Humboldt Foundation for funding.

\section*{Code Availability}
The code used to reproduce the results in this paper can be found at \cite{githubrepomhn}.

\paragraph*{Declaration of AI usage} ChatGPT 5.6 was used for assistance with coding and text editing. All outputs were reviewed and verified by the authors.

\putbib[references]

\end{bibunit}


\onecolumngrid


\setcounter{section}{0}
\setcounter{subsection}{0}
\setcounter{subsubsection}{0}

\setcounter{equation}{0}
\setcounter{figure}{0}
\setcounter{table}{0}

\renewcommand{\theequation}{S\arabic{equation}}
\renewcommand{\thefigure}{S\arabic{figure}}
\renewcommand{\thetable}{S\arabic{table}}

\renewcommand{\thesection}{\Roman{section}}
\renewcommand{\thesubsection}{\Alph{subsection}}

\setcounter{page}{1}




\begin{center}

{\large\bfseries
Supplementary Information for\\[0.3em]
A Bifurcation Theory for the Equilibria of Modern Hopfield Networks
}

\vspace{1em}

Vincenzo Maria Schimmenti$^{1}$ and
Matteo Ciarchi$^{1,2,3}$

\vspace{0.5em}

{\small
$^{1}$Max-Planck-Institute for the Physics of Complex Systems,
Noethnitzer Str. 38, 01187 Dresden, Germany\\
$^{2}$Ludwig-Maximilians-Universit\"at M\"unchen,
Arnold-Sommerfeld-Center for Theoretical Physics,
Theresienstr. 37, 80333 M\"unchen, Germany\\
$^{3}$Gurdon Institute, University of Cambridge,
Cambridge CB2 1QN, UK
}

\end{center}

\vspace{1em}


\begin{bibunit}[apsrev4-2]

\section{Useful identities and notation}
In the following, we will employ the operation "softmax" as a map from vectors in $\mathbb{R}^K$ to points in the $K$-dimensional simplex $\Delta_K$
\begin{equation}
    \boldsymbol{p}(\boldsymbol{\theta}) = \operatorname{softmax}(\boldsymbol{\theta)} 
\end{equation}
Component-wise:
\begin{equation}
    p_\mu(\boldsymbol{\theta}) = \frac{e^{\theta_\mu}}{\sum_{\nu=1}^K e^{\theta_\nu}}
\end{equation}
We will also employ the Fisher matrix of a point on the simplex:
\begin{equation}
    \boldsymbol{F}(\boldsymbol{p}) = \operatorname{diag}({\boldsymbol{p}}) - \boldsymbol{p}  \boldsymbol{p}^\intercal
\end{equation}
which is also the derivative of $\boldsymbol{p}(\boldsymbol{\theta})$ w.r.t. $\boldsymbol{\theta}$:
\begin{equation}
    \frac{\partial p_\mu (\boldsymbol{\theta)} }{\partial \theta_\nu} = F_{\mu \nu}(\boldsymbol{p}(\boldsymbol{\theta}))
\end{equation}
The Hessian reads:
\begin{equation}
    \frac{\partial^2 p_\mu}{\partial \theta_\rho  \partial \theta_\nu} = \sum_{\sigma} \frac{\partial F_{\mu \nu}}{\partial p_\sigma} \frac{\partial p_\sigma}{\partial \theta_\rho} = \sum_{\sigma} (\delta_{\mu \sigma} \delta_{\nu \sigma} - \delta_{\mu \sigma} p_\nu - p_\mu \delta_{\nu \sigma}) F_{\sigma \rho} = \delta_{\mu \nu} F_{\mu \rho} - F_{\mu \rho} p_\nu - p_\mu F_{\nu \rho}
\end{equation}

Another useful identity is the convex-dual form of a log-sum-exp term:
\begin{equation}
    \log \left(\sum_{\mu=1}^K e^{\theta_\mu}\right) = \sup_{\boldsymbol{w} \in \Delta_K} \left(\sum_{\mu=1}^K \theta_\mu  w_\mu + H(\boldsymbol{w}) \right)
\end{equation}
where $\boldsymbol{w}$ is a point in the $K$-dimensional simplex and $H(\boldsymbol{w})=- \sum_{\mu=1}^K w_\mu \log w_\mu$ is its Shannon entropy. Finally, we will make use of the softmax expansion around a point $\boldsymbol{\theta}_0$. 
This yields, for a generic fixed-point of the softmax map:
\begin{equation}
\boldsymbol{w}(\boldsymbol{\theta}) = \boldsymbol{w}_0 + \boldsymbol{F}_0\,\delta\boldsymbol{\theta}
+ \frac{1}{2}\left[\delta\boldsymbol{\theta} - \left(\boldsymbol{w}_0^\intercal \delta\boldsymbol{\theta}\right)\boldsymbol{1}\right] \odot \left(\boldsymbol{F}_0\delta\boldsymbol{\theta}\right)
- \frac{1}{2}\boldsymbol{w}_0\left(\delta\boldsymbol{\theta}^\intercal \boldsymbol{F}_0 \delta\boldsymbol{\theta}\right)
+ O\!\left(\|\delta\boldsymbol{\theta}\|^3\right).
\end{equation}

\section{Deterministic dynamics}
In this section, we analyze the deterministic dynamics of the gradient-ascent evolution of an MHN in the direct space of the features $\boldsymbol{x}$. We focus on the general formulation of a Modern Hopfield Network with log-sum-exp potential. The energy function associated is:
\begin{equation}
E(\boldsymbol{x})=\frac{1}{2}||\boldsymbol{x}||^2-\frac{1}{\beta}\log\left(\sum_{\mu=1}^K \exp\left(\beta(\boldsymbol{\xi}_\mu\cdot\boldsymbol{x}+a_\mu)\right)\right).
\end{equation}
Taking $a_\mu = - \frac{1}{2} ||\boldsymbol{\xi}_\mu||^2$ yields the "Euclidean" version discussed in the main text. The gradient descent dynamics associated with this energy reads:
\begin{equation}
\dot{\boldsymbol{x}}=\boldsymbol{\Xi}^\intercal \operatorname{softmax}\left(\beta (\boldsymbol{\Xi} \boldsymbol{x} + \boldsymbol{a}) \right)-\boldsymbol{x}.
\end{equation}
where we introduced the pattern matrix $\boldsymbol{\Xi}$ with components $\Xi_{\mu,i}=(\boldsymbol{\xi}_\mu)_i$. As a shorthand, we will denote the softmax as $\boldsymbol{w}(\boldsymbol{x})$. The Hessian of the energy reads:
\begin{equation}
    \mathcal{H}(\boldsymbol{x}) = \boldsymbol{I} - \beta \boldsymbol{\Xi}^\intercal \boldsymbol{F}(\boldsymbol{w}(\boldsymbol{x})) \boldsymbol{\Xi} 
\end{equation}
A fixed point $\boldsymbol{x}^*$ is stable if the smallest eigenvalue of the Hessian is positive. Equivalently:
\begin{equation}
    \beta < \frac{1}{ \lambda_{\rm max} \left(\boldsymbol{\Xi}^\intercal \boldsymbol{F}(\boldsymbol{w}(\boldsymbol{x}^*)) \boldsymbol{\Xi}\right)}
\end{equation}
when $\beta=0$, the only fixed point is the uniform one, namely $\boldsymbol{x}^* = \frac{1}{K} \sum_{\mu=1}^K \boldsymbol{\xi}_\mu \equiv \boldsymbol{\bar{\xi}}$. Equivalently, the corresponding fixed point in the simplex space reads $\boldsymbol{w}(\boldsymbol{x}^*) = \frac{1}{K} \boldsymbol{1}$. This implies that the critical $\beta$ for the uniform state is simply:
\begin{equation}
    \beta_c = \frac{K}{ \lambda_{\rm max} \left(\boldsymbol{\Xi}^\intercal \left( \boldsymbol{I} - \frac{1}{K} \boldsymbol{1} \boldsymbol{1}^\intercal \right)\boldsymbol{\Xi}\right)}
\end{equation}
However, in general, the uniform fixed point is not a stable fixed point at all $\beta$ values. We do not study further the matter of stability, as it will be more easily formulated into the dual approach.
\section{Dual deterministic dynamics}
The same analysis on the deterministic dynamics can be obtained for the convex-dual formulation. A useful form for the analysis of the MHN fixed points leverages the convex dual of the log-sum-exp term. 
\begin{equation}
    \min_{\boldsymbol{x}} E(\boldsymbol{x}) = \min_{\boldsymbol{x}} \left[ \frac{1}{2} ||\boldsymbol{x}||^2 - \frac{1}{\beta} \sup_{\boldsymbol{w}} \left( H(\boldsymbol{w}) + \beta \boldsymbol{w}^\intercal (\boldsymbol{\Xi} \boldsymbol{x} + \boldsymbol{a})\right) \right]
\end{equation}
Moving the sup operation outside the minus sign gives the equivalent condition:
\begin{equation}
    \min_{\boldsymbol{x}} E(\boldsymbol{x}) = \min_{\boldsymbol{x}}  \inf_{\boldsymbol{w}} \left[ \frac{1}{2} ||\boldsymbol{x}||^2 - \frac{1}{\beta} H(\boldsymbol{w}) -\boldsymbol{w}^\intercal (\boldsymbol{\Xi} \boldsymbol{x} + \boldsymbol{a}) \right]
    \label{eq:convexdualminreKaio}
\end{equation}
Minimizing over $\boldsymbol{x}$ gives:
\begin{equation}
    \boldsymbol{x}(\boldsymbol{w}) = \boldsymbol{\Xi}^\intercal \boldsymbol{w}
\end{equation}
Inserting the minimizer value of $\boldsymbol{x}$ back into Eq.~\eqref{eq:convexdualminreKaio}:
\begin{equation}
    \inf_{\boldsymbol{w}} \left[ - \frac{1}{\beta} H(\boldsymbol{w}) - \boldsymbol{w}^\intercal \boldsymbol{a} - \frac{1}{2} \boldsymbol{w}^\intercal \boldsymbol{G} \boldsymbol{w} \right]
\end{equation}
where $\boldsymbol{G} = \boldsymbol{\Xi} \boldsymbol{\Xi}^\intercal$ is the Gram matrix of the patterns. The infimum condition is equivalent to:
\begin{equation}
    \sup_{\boldsymbol{w}} \left[\frac{1}{\beta} H(\boldsymbol{w}) + \boldsymbol{w}^\intercal \boldsymbol{a} + \frac{1}{2} \boldsymbol{w}^\intercal \boldsymbol{G} \boldsymbol{w} \right]
\end{equation}
The same result can be obtained from a Legendre transform of the MHN energy. This thus defines the functional $\Phi(\boldsymbol{w})=\left[\frac{1}{\beta} H(\boldsymbol{w}) + \boldsymbol{w}^\intercal \boldsymbol{a} + \frac{1}{2} \boldsymbol{w}^\intercal \boldsymbol{G} \boldsymbol{w} \right]$ whose maximization gives the fixed-point of MHNs. The stability condition is also retained: for an interior stationary point, the Hessian of $\Phi(\boldsymbol{w})$ reads:
\begin{equation}
    \nabla^2\Phi = G-\frac{1}{\beta}\text{diag}(1/\boldsymbol{w}).
\end{equation}
This implies that the negative definiteness of the Hessian is equivalent to
\begin{equation}
    \beta \lambda_{\text{max}}(F^{1/2}GF^{1/2})<1,
\end{equation}
as obtained above.

The dynamical equation for $\boldsymbol{w}$ can be obtained using its natural gradient, but since we are interested in the fixed point, we can employ simple gradient descent:
\begin{equation}
    \dot{\boldsymbol{w}} = \operatorname{softmax}(\beta (\boldsymbol{G} \boldsymbol{w} + \boldsymbol{a})) - \boldsymbol{w},
\end{equation}
which respects the constraint $\sum_{\mu=1}^K w_\mu = 1$ and positivity. The Jacobian of this map reads:
\begin{equation}
    \boldsymbol{J}(\boldsymbol{w}) = \beta \boldsymbol{F}(\boldsymbol{p}(\boldsymbol{w})) \boldsymbol{G} - \boldsymbol{I},
\end{equation}
with $\boldsymbol{p}(\boldsymbol{w}) = \operatorname{softmax}(\beta (\boldsymbol{G} \boldsymbol{w} + \boldsymbol{a}))$. At a fixed point $\boldsymbol{w} = \boldsymbol{p}(\boldsymbol{w})$, hence:
\begin{equation}
    \boldsymbol{J}(\boldsymbol{w}) = \beta \boldsymbol{F}(\boldsymbol{w}) \boldsymbol{G} - \boldsymbol{I}.
\end{equation}

It is worth mentioning that, while our gradient ascent correctly identifies the fixed point, if the true dynamics on the energy landscape is of interest, the correct ascent is the natural gradient one, namely:
\begin{equation}
   \dot{\boldsymbol{w}} = \boldsymbol{F}(\boldsymbol{w}) \nabla_{\boldsymbol{w}} \Phi.
\end{equation}
This comes from the gradient of the convex-dual functional:
\begin{equation}
    \nabla \Phi = \boldsymbol{a} + \boldsymbol{G} \boldsymbol{w} - \frac{1}{\beta} (\log \boldsymbol{w}+\boldsymbol{1}),
\end{equation}
where the log is intended to be component-wise. Thus:
\begin{equation}
    \dot{w} = \left(\operatorname{diag}(\boldsymbol{w}) - \boldsymbol{w} \boldsymbol{w}^\intercal  \right) \left( \boldsymbol{a} + \boldsymbol{G} \boldsymbol{w} - \frac{1}{\beta} (\log \boldsymbol{w}+\boldsymbol{1})\right).
\end{equation}
This can be interpreted as replicator dynamics:
\begin{equation}
    \dot{\boldsymbol{w}} = \boldsymbol{w} \odot \left( \boldsymbol{a} + \boldsymbol{G} \boldsymbol{w} - \frac{1}{\beta} (\log \boldsymbol{w} + 1)  - \langle \boldsymbol{a} + \boldsymbol{G} \boldsymbol{w} - \frac{1}{\beta} (\log \boldsymbol{w} + 1) \rangle_{\boldsymbol{w}} \right),
\end{equation}
which basically makes the simplex weights navigate proportionally (per component) to the difference between the gradient and its expected value w.r.t. the weights themselves. When extended to stochastic systems, this leads to Wright-Fisher diffusion \cite{aurell2019multilocuswrightfishermodelmutation}.
\subsection{Expansion of convex-dual functional around a fixed point}
We now expand the convex-dual free energy around a fixed point $\boldsymbol{w}_0$ as $\boldsymbol{w} = \boldsymbol{w}_0 + \epsilon \boldsymbol{\hat{t}}$, where $\boldsymbol{\hat{t}}$ is the eigenvector associated with the largest eigenvalue of the projected Gram matrix:
\begin{equation}
\boldsymbol{F}(\boldsymbol{w}_0)\boldsymbol{G}\boldsymbol{\hat{t}} = \frac{1}{\beta_c}\boldsymbol{\hat{t}}, \qquad \boldsymbol{\hat{t}}^\intercal\boldsymbol{\hat{t}}=1.
\end{equation}

The logarithm in the entropy is, up to third order:
\begin{equation}
\log(w_{\mu,0}+\epsilon\hat{t}_\mu) \approx \log(w_{\mu,0}) + \epsilon\frac{\hat{t}_\mu}{w_{\mu,0}} - \frac{\epsilon^2}{2}\frac{\hat{t}_\mu^2}{w_{\mu,0}^2} + \frac{\epsilon^3}{3}\frac{\hat{t}_\mu^3}{w_{\mu,0}^3}.
\end{equation}

This gives an expansion of the entropy at the same order:
\begin{equation}
    \begin{aligned}
        H(\boldsymbol{w}) = & -\sum_{\mu=1}^K(w_{\mu,0}+\epsilon\hat{t}_\mu)\log(w_{\mu,0}+\epsilon\hat{t}_\mu)
\approx H(\boldsymbol{w}_0) - \epsilon\boldsymbol{1}^\intercal\boldsymbol{\hat{t}} - \epsilon\sum_{\mu=1}^K\hat{t}_\mu\log(w_{\mu,0})\\
& - \frac{\epsilon^2}{2}\sum_{\mu=1}^K\frac{\hat{t}_\mu^2}{w_{\mu,0}} + \frac{\epsilon^3}{6}\sum_{\mu=1}^K\frac{\hat{t}_\mu^3}{w_{\mu,0}^2} + O(\epsilon^4).
    \end{aligned}
\end{equation}

The quadratic interaction is expanded up to second order:
\begin{equation}
\frac{1}{2}\boldsymbol{w}^\intercal\boldsymbol{G}\boldsymbol{w} \approx \frac{1}{2}\boldsymbol{w}_0^\intercal\boldsymbol{G}\boldsymbol{w}_0 + \epsilon \boldsymbol{\hat{t}}^\intercal\boldsymbol{G}\boldsymbol{w}_0 + \frac{\epsilon^2}{2}\boldsymbol{\hat{t}}^\intercal\boldsymbol{G}\boldsymbol{\hat{t}},
\end{equation}
where we used the fact that $\boldsymbol{G}$ is symmetric, and
\begin{equation}
\boldsymbol{w}^\intercal\boldsymbol{a} = \boldsymbol{w}_0^\intercal\boldsymbol{a} + \epsilon \boldsymbol{\hat{t}}^\intercal\boldsymbol{a}.
\end{equation}

Now we notice the following property for the sum of the components of the dominant eigenvector:
\begin{equation}
\sum_{\mu=1}^K\hat{t}_\mu = \beta_c\sum_{\mu,\rho,\nu=1}^K\left(w_{\mu,0}\delta_{\mu\rho}-w_{\mu,0}w_{\rho,0}\right)G_{\rho\nu}\hat{t}_\nu = \beta_c\sum_{\rho,\nu=1}^K(w_{\rho,0}-w_{\rho,0})G_{\rho\nu}\hat{t}_\nu = 0,
\end{equation}
which, given the tangency of $\boldsymbol{\hat{t}}$ to the simplex, always holds. This removes one of the linear terms in the expanded entropy. The expansion of the convex-dual potential thus reads:
\begin{equation}
\begin{aligned}
    \Phi \approx & \frac{1}{2}\boldsymbol{w}_0^\intercal\boldsymbol{G}\boldsymbol{w}_0 + \boldsymbol{w}_0^\intercal\boldsymbol{a}
+\left(\boldsymbol{\hat{t}}^\intercal\boldsymbol{G}\boldsymbol{w}_0+\boldsymbol{\hat{t}}^\intercal\boldsymbol{a}\right)\epsilon
+ \frac{\epsilon^2}{2}\boldsymbol{\hat{t}} \\ & ^\intercal\boldsymbol{G}\boldsymbol{\hat{t}}
+ \frac{1}{\beta}\left[H(\boldsymbol{w}_0)-\epsilon\sum_{\mu=1}^K\hat{t}_\mu\log(w_{\mu,0})-\frac{\epsilon^2}{2}\sum_{\mu=1}^K\frac{\hat{t}_\mu^2}{w_{\mu,0}}+\frac{\epsilon^3}{6}\sum_{\mu=1}^K\frac{\hat{t}_\mu^3}{w_{\mu,0}^2}\right] + O(\epsilon^4).
\end{aligned}
\end{equation}
We can further simplify the quadratic term:
\begin{equation}
    \frac{1}{2}\boldsymbol{\hat{t}}^\intercal\boldsymbol{G}\boldsymbol{\hat{t}} = \frac{1}{2\beta_c} \sum_{\mu=1}^K \frac{\hat{t}_\mu^2}{w_{\mu,0}} 
\end{equation}
yielding:
\begin{equation}
    \Phi - \Phi_0 \approx \textrm{const} + \boldsymbol{\hat{t}}^\intercal (\boldsymbol{G} \boldsymbol{w}_0 + \boldsymbol{a} ) \epsilon -\frac{\epsilon}{\beta } \sum_{\mu=1}^K \log(w_{\mu,0}) \hat{t}_\mu + \frac{\epsilon^2}{2} \left(\frac{1}{\beta_c} - \frac{1}{\beta}\right) \sum_{\mu=1}^K \frac{\hat{t}_\mu^2}{w_{\mu,0}} + \frac{1}{\beta}\frac{\epsilon^3}{6}\sum_{\mu=1}^K\frac{\hat{t}_\mu^3}{w_{\mu,0}^2} + O(\epsilon^4)
\end{equation}
Since $\beta_0 \leq \beta \leq \beta_c$, the coefficient of the second-order term in $\epsilon$ is negative.

Evaluating the expanded potential at the uniform fixed point $w_{\mu,0}=1/K$ we obtain:
\begin{equation}
\Phi \approx \frac{1}{2}\boldsymbol{w}_0^\intercal\boldsymbol{G}\boldsymbol{w}_0 + \boldsymbol{w}_0^\intercal\boldsymbol{a}_\mu +
\left(\boldsymbol{\hat{t}}^\intercal\boldsymbol{\Xi}\boldsymbol{\bar{\xi}}+\boldsymbol{\hat{t}}^\intercal\boldsymbol{a}\right)\epsilon
+ \frac{ K \epsilon^2}{2\beta_c} 
+ \frac{1}{\beta}\left[H(\boldsymbol{w}_0)-\frac{K\epsilon^2}{2}+\frac{K^2\epsilon^3}{6}\sum_{\mu=1}^K\hat{t}_\mu^3\right] + O(\epsilon^4),
  \end{equation}
where $\boldsymbol{\bar{\xi}}$ denotes the average pattern. For the uniform mode, the eigenvector $\hat{t}$ has a clean interpretation. From its definition, the Fisher matrix is proportional to a projector onto the vector of ones i.e. $\boldsymbol{F}(\boldsymbol{w}_0) = \frac{1}{K} \left(\boldsymbol{I} - \frac{1}{K} \boldsymbol{1} \boldsymbol{1}^\intercal\right) \equiv \frac{1}{K} \boldsymbol{P}$. Given that $\boldsymbol{P} \boldsymbol{\hat{t}} = \boldsymbol{\hat{t}}$, we obtain:
\begin{equation}
        \frac{1}{K} \boldsymbol{P} \boldsymbol{G} \boldsymbol{\hat{t}} = \frac{1}{\beta_c} \boldsymbol{\hat{t}} \implies
        \boldsymbol{P} \boldsymbol{G} \boldsymbol{P} \boldsymbol{\hat{t}} = \frac{K}{\beta_c} \boldsymbol{\hat{t}} \implies
        \boldsymbol{P} \boldsymbol{\Xi} \boldsymbol{\Xi}^\intercal \boldsymbol{P} \boldsymbol{\hat{t}} = \frac{K}{\beta_c} \boldsymbol{\hat{t}}
\end{equation}
Defining the centered pattern matrix as $\boldsymbol{\Xi}_c = \boldsymbol{P} \boldsymbol{\Xi}$, we get:
\begin{equation}
    \boldsymbol{\Xi}_c \boldsymbol{\Xi}_c^\intercal \boldsymbol{\hat{t}} = \frac{K}{\beta_c} \boldsymbol{\hat{t}}
\end{equation}
Decomposing $\boldsymbol{\Xi}_c$ via an SVD $\boldsymbol{\Xi}_c = \boldsymbol{U}_c \boldsymbol{\Sigma}_c \boldsymbol{V}_c^\intercal$ we immediately find that $\boldsymbol{\hat{t}} = \boldsymbol{u}_{\rm max}$ and $\beta_c = \frac{K}{\sigma_{\rm max}^2}$, where $\sigma_{\rm max}$ is the largest principal value and $\boldsymbol{u}_{\rm max}$ is associated eigenvector. We also have:

\begin{equation}
    \boldsymbol{\Xi}_c \boldsymbol{v}_{\rm max} = \sigma_{\rm max} \boldsymbol{u}_{\rm max}
\end{equation}
with $\boldsymbol{v}_{\rm max}$ the first principal component. Consequently, we find that the eigenvector $\boldsymbol{\hat{t}}$ is simply the projection of the centered pattern vector onto the first principal component:
\begin{equation}
    \hat{t}_\mu = \frac{(\boldsymbol{\xi}_\mu - \boldsymbol{\bar{\xi}}) \cdot \boldsymbol{v}_{\rm max}}{\sigma_{\rm max}}
\end{equation}
In this language, the pattern-dependent linear term becomes:
\begin{equation}
    \boldsymbol{\hat{t}}^\intercal \boldsymbol{\Xi} \boldsymbol{\bar{\xi}} =  \sigma_{\rm max} \boldsymbol{v}_{\rm max} \cdot \boldsymbol{\bar{\xi}} = \sqrt{\frac{K}{\beta_c}} \boldsymbol{v}_{\rm max} \cdot \boldsymbol{\bar{\xi}}
\end{equation}
This immediately tells us that the third-order term measures the skewness of the data and the fourth-order one (if we were to keep it), its kurtosis. We are ready to classify the bifurcation type for the uniform fixed point:
\begin{enumerate}
    \item \textbf{No bifurcation/imperfect bifurcation}: linear term is non-zero. The uniform fixed point becomes immediately unstable, as the position of the unique fixed point at low $\beta$ simply shifts. When no bias is present and for centered patterns, this term vanishes. Moreover, an ad-hoc choice of biases, namely $a_\mu = - \boldsymbol{\xi}_\mu \cdot \boldsymbol{\bar{\xi}}$, also makes this term vanish. The latter option effectively recenters the patterns.
    \item \textbf{Stationary point}: when $\beta < \beta_c$, the uniform fixed point is simply a stable stationary point.
    \item \textbf{Transcritical}: no linear term and $\beta=\beta_c$. If the cubic term is non-zero (skewed case), the bifurcation is transcritical/asymmetric. If $\epsilon > 0$, patterns on the positive sign of $\boldsymbol{v}_{\rm max}$ get additional weight. 
    \item \textbf{Pitchfork}: no linear term, $\beta=\beta_c$ and no skewness. In this case, we need to keep the fourth-order term in the expansion, and it yields a pitchfork bifurcation. 
\end{enumerate}
It is worth discussing the result for the Euclidean case, where the linear term acquires a particularly meaningful form. The product of a pattern with the average reads:
\begin{equation}
    \boldsymbol{\xi}_\mu \cdot \boldsymbol{\bar{\xi}} - \frac{1}{2} ||\boldsymbol{\xi}_\mu||^2 = (\boldsymbol{\bar{\xi}} + \delta \boldsymbol{\xi}_\mu) \cdot \boldsymbol{\bar{\xi}} - \frac{1}{2} ||\boldsymbol{\bar{\xi}} + \delta \boldsymbol{\xi}_\mu||^2 = \frac{1}{2} ||\boldsymbol{\bar{\xi}}||^2 - \frac{1}{2} || \delta \boldsymbol{\xi}_\mu||^2.
\end{equation}
As a consequence:
\begin{equation}
    \sum_{\mu=1}^K \hat{t}_\mu (\boldsymbol{\xi}_\mu \cdot \boldsymbol{\bar{\xi}} - \frac{1}{2} ||\boldsymbol{\xi}_\mu||^2) = - \frac{1}{2} \sum_{\mu=1}^K \hat{t}_\mu || \boldsymbol{\xi}_\mu - \boldsymbol{\bar{\xi}}||^2.
\end{equation}
For centered and normalized patterns (i.e., patterns on an $N$-dimensional sphere), this term vanishes, and we can effectively have a bifurcation in the Euclidean case. 

\section{Toy models}

We now consider some examples of simple models where the study of the hierarchy of bifurcations is analytically possible.

\subsection{Gram matrix proportional to the identity}

We start by considering the simple case of a Gram matrix proportional to the identity matrix, $G=gI$, on the space of patterns. The convex-dual free energy reads:
\begin{equation}
    F(\boldsymbol{w}) = \frac{g}{2} || \boldsymbol{w}||^2 + \boldsymbol{a}^\intercal \boldsymbol{w} + \frac{1}{\beta}  H(\boldsymbol{w})
\end{equation}
When $\boldsymbol{a} = \boldsymbol{0}$, due to the symmetry of patterns, we can consider solutions of the type $w_\mu =\frac1m$  for a subset of $m$ patterns. The free energy is:
\begin{equation}
    F_m = \frac{\beta g}{2m} + \log m,
\end{equation}
where $g$ is the norm of the patterns. Given the convexity of the free energy, the only maximisers are $m=1$ and $m=K$. The free energy matches at:
\begin{equation}
    F_1(\beta_c) = F_{K}(\beta_c)
\end{equation}
implying:
\begin{equation}
    \beta_c = \frac{2}{g} \frac{K}{K-1} \log K
\end{equation}
A similar situation occurs for the Euclidean case, where $a_\mu = - \frac{1}{2} ||\boldsymbol{\xi}_\mu||^2 = -\frac{g}{2}$ is constant and thus does not contribute to the maximization. 
\subsection{1-step hierarchical patterns }
We now consider the scenario where the patterns have a Gram structure divided into $M$ equal blocks, each of size $\alpha=K/M$. Denoting $b(\mu)$ the block index of pattern $\mu$, the Gram matrix reads:
\begin{equation}
    G_{\mu \nu} =  g \rho_0 + g (\rho_1 -\rho_0) \delta_{b(\mu) b(\nu)} +  g(1-\rho_1)\delta_{\mu \nu}.
\end{equation}
For $\rho_1 = \rho_0$, there is no block structure, and we recover the case of equally correlated patterns. We fix, for simplicity, $g=1$. We surmise three possible fixed points. The uniform one with $w_\mu = 1/K$ with free energy:
\begin{equation}
    F_K = \log K + \frac{\beta}{2 K} (1 + (\alpha-1) \rho_1 + (K-\alpha) \rho_0) 
\end{equation}
The other extreme solution is localized in a single pattern, whose free energy reads:
\begin{equation}
    F_1 = \frac{\beta}{2} G_{\mu \mu} = \frac{\beta}{2}
\end{equation}
Finally, an intermediate case can arise, which is equally uniform inside a block:
\begin{equation}
    F_M = \log \alpha + \frac{\beta}{2 \alpha} (1 + (\alpha-1) \rho_1)  
\end{equation}
We can either go from a uniform to a state localized to a single pattern by bifurcating via a state localized on a single block, or directly. As before, since all the patterns have the same norm, the "uniform prior" version gives the same maximization conditions. 

\section{Continuum limit of a hierarchical Gram matrix}
\label{sec:hierarchical_continuum}

In this section, we derive the continuum hierarchical free energy introduced in the main text. 
We start from a finite hierarchy of nested blocks and subsequently take the limit of an infinite 
number of hierarchical levels.

\subsection{Finite hierarchical construction}

Consider $K$ patterns organized into a sequence of nested partitions
\begin{equation}
    \mathcal{P}_0 \supset \mathcal{P}_1 \supset \cdots \supset \mathcal{P}_L ,
\end{equation}
where each block at level $l$ contains $N_l$ patterns. We assume a regular hierarchy,
\begin{equation}
    N_0 = K,
    \qquad
    N_0 > N_1 > \cdots > N_L,
\end{equation}
and that $N_l/N_{l+1}$ is an integer. The label $b_l(\mu)$ denotes the block containing
pattern $\mu$ at hierarchical level $l$.

We associate with every level an overlap $q_l$, with
\begin{equation}
    q_0 \leq q_1 \leq \cdots \leq q_L ,
\end{equation}
so that patterns sharing a finer block have a larger overlap. A convenient representation of
the corresponding hierarchical Gram matrix is
\begin{equation}
    G_{\mu\nu}
    =
    q_0
    +
    \sum_{r=1}^{L}
    \Delta q_r\,
    \delta_{b_r(\mu),b_r(\nu)},
    \qquad
    \Delta q_r \equiv q_r-q_{r-1}.
    \label{eq:hierarchical_gram_discrete}
\end{equation}
If $N_L=1$ and $q_L=1$, this construction also fixes the diagonal entries to unity.

We now consider a candidate fixed point uniformly localized on one block $B_l$ of level $l$,
\begin{equation}
    w_\mu^{(l)}
    =
    \begin{cases}
        N_l^{-1}, & \mu\in B_l,\\
        0, & \mu\notin B_l .
    \end{cases}
    \label{eq:hierarchical_trial_state}
\end{equation}
Its entropy is simply
\begin{equation}
    H\!\left(\boldsymbol{w}^{(l)}\right)
    =
    \log N_l.
    \label{eq:hierarchical_entropy}
\end{equation}

The quadratic contribution to the convex-dual functional is
\begin{equation}
    \left(\boldsymbol{w}^{(l)}\right)^\top
    \boldsymbol{G}
    \boldsymbol{w}^{(l)}
    =
    \frac{1}{N_l^2}
    \sum_{\mu,\nu\in B_l}
    G_{\mu\nu}.
    \label{eq:hierarchical_energy_average}
\end{equation}
To evaluate this quantity, notice that for every $r\leq l$, two patterns belonging to $B_l$
necessarily belong to the same block at level $r$. These contributions therefore sum to $q_l$.

For a finer level $r>l$, the block $B_l$ contains $N_l/N_r$ sub-blocks of size $N_r$.
The fraction of ordered pairs $(\mu,\nu)\in B_l\times B_l$ that belong to the same
$r$-level block is therefore
\begin{equation}
    \frac{
        (N_l/N_r)N_r^2
    }{
        N_l^2
    }
    =
    \frac{N_r}{N_l}.
\end{equation}
It follows that
\begin{equation}
    \left(\boldsymbol{w}^{(l)}\right)^\top
    \boldsymbol{G}
    \boldsymbol{w}^{(l)}
    =
    q_l
    +
    \sum_{r=l+1}^{L}
    \frac{N_r}{N_l}
    \Delta q_r .
    \label{eq:hierarchical_energy_discrete}
\end{equation}

Consequently, the value of the convex-dual objective for a state localized at level $l$ is
\begin{equation}
    \Phi_l
    =
    \frac{1}{\beta}\log N_l
    +
    \frac{1}{2}
    \left[
        q_l
        +
        \sum_{r=l+1}^{L}
        \frac{N_r}{N_l}
        \left(q_r-q_{r-1}\right)
    \right].
    \label{eq:hierarchical_free_energy_discrete}
\end{equation}
The selected hierarchical level is the one maximizing $\Phi_l$.

Equation~\eqref{eq:hierarchical_free_energy_discrete} makes explicit the competition responsible
for the hierarchical bifurcation cascade. The entropy favors large blocks, while the quadratic
interaction favors smaller blocks containing more strongly correlated patterns. Increasing $\beta$
reduces the relative importance of the entropy and therefore progressively favors finer levels of
the hierarchy.

For two levels $l$ and $m$, their coexistence inverse temperature is determined by
$\Phi_l=\Phi_m$. Writing
\begin{equation}
    E_l
    \equiv
    \frac{1}{2}
    \left[
        q_l
        +
        \sum_{r=l+1}^{L}
        \frac{N_r}{N_l}
        \left(q_r-q_{r-1}\right)
    \right],
\end{equation}
one obtains
\begin{equation}
    \frac{1}{\beta_{l,m}}
    =
    \frac{E_m-E_l}
    {\log N_l-\log N_m}.
    \label{eq:hierarchical_crossing_discrete}
\end{equation}
Thus, depending on the shape of the sequence $(\log N_l,E_l)$, the optimal state can either
move through consecutive hierarchical levels or jump directly across several levels.

\subsection{Continuum limit}

We now consider the limit $L\rightarrow\infty$ and introduce the continuous hierarchical variable
\begin{equation}
    s=\frac{l}{L}\in[0,1].
\end{equation}
The block size and overlap become smooth functions,
\begin{equation}
    N_l \longrightarrow N(s),
    \qquad
    q_l \longrightarrow q(s),
\end{equation}
with
\begin{equation}
    N'(s)<0,
    \qquad
    q'(s)>0,
\end{equation}
corresponding respectively to decreasing block size and increasing pattern correlation as one moves
toward finer hierarchical levels.

In the continuum limit,
\begin{equation}
    q_r-q_{r-1}
    \longrightarrow
    q'(u)\,du,
\end{equation}
and the sum in Eq.~\eqref{eq:hierarchical_free_energy_discrete} becomes
\begin{equation}
    \sum_{r=l+1}^{L}
    \frac{N_r}{N_l}
    \left(q_r-q_{r-1}\right)
    \longrightarrow
    \int_s^1
    \frac{N(u)}{N(s)}
    q'(u)\,du.
\end{equation}
The hierarchical free energy therefore becomes
\begin{equation}
    \boxed{
    \mathcal{F}(s)
    =
    \frac{1}{\beta}\log N(s)
    +
    \frac{1}{2}
    \left[
        q(s)
        +
        \int_s^1
        \frac{N(u)}{N(s)}
        q'(u)\,du
    \right]
    }
    \label{eq:hierarchical_free_energy_continuum}
\end{equation}
which is the expression of the main text.

It is useful to define
\begin{equation}
    I(s)
    \equiv
    \int_s^1
    \frac{N(u)}{N(s)}
    q'(u)\,du.
    \label{eq:I_hierarchy}
\end{equation}
Since
\begin{equation}
    I(s)
    =
    \frac{1}{N(s)}
    \int_s^1
    N(u)q'(u)\,du,
\end{equation}
its derivative is
\begin{align}
    I'(s)
    &=
    -\frac{N'(s)}{N(s)^2}
    \int_s^1 N(u)q'(u)\,du
    -q'(s)
    \nonumber\\
    &=
    -\frac{N'(s)}{N(s)}I(s)-q'(s).
    \label{eq:Iprime_hierarchy}
\end{align}
Using Eq.~\eqref{eq:Iprime_hierarchy}, the derivative of
Eq.~\eqref{eq:hierarchical_free_energy_continuum} becomes
\begin{align}
    \mathcal{F}'(s)
    &=
    \frac{1}{\beta}\frac{N'(s)}{N(s)}
    +
    \frac{1}{2}\left[q'(s)+I'(s)\right]
    \nonumber\\
    &=
    \frac{N'(s)}{N(s)}
    \left[
        \frac{1}{\beta}
        -
        \frac{I(s)}{2}
    \right].
    \label{eq:hierarchical_free_energy_derivative}
\end{align}
This is the derivative obtained in the main text.

An interior optimum $s^*$ therefore satisfies
\begin{equation}
    I(s^*)
    =
    \frac{2}{\beta}.
    \label{eq:hierarchical_stationarity}
\end{equation}
If
\begin{equation}
    I'(s)<0
    \qquad
    \forall s\in[0,1],
    \label{eq:I_monotonic_condition}
\end{equation}
then Eq.~\eqref{eq:hierarchical_stationarity} has at most one solution for any fixed $\beta$.
Moreover, differentiating Eq.~\eqref{eq:hierarchical_stationarity} with respect to $\beta$ gives
\begin{equation}
    I'(s^*)\frac{ds^*}{d\beta}
    =
    -\frac{2}{\beta^2},
\end{equation}
and therefore
\begin{equation}
    \frac{ds^*}{d\beta}
    =
    -\frac{2}{\beta^2 I'(s^*)}.
    \label{eq:hierarchical_s_beta}
\end{equation}
Under condition~\eqref{eq:I_monotonic_condition}, this implies
\begin{equation}
    \frac{ds^*}{d\beta}>0.
\end{equation}
Hence, increasing $\beta$ continuously selects finer hierarchical levels.

To make the geometric interpretation more explicit, define the effective energetic contribution
\begin{equation}
    E(s)
    =
    \frac{1}{2}\left[q(s)+I(s)\right].
    \label{eq:hierarchical_Es}
\end{equation}
Then
\begin{equation}
    \mathcal{F}(s)
    =
    E(s)
    +
    \frac{1}{\beta}\log N(s).
    \label{eq:hierarchical_F_energy_entropy}
\end{equation}
From Eq.~\eqref{eq:Iprime_hierarchy},
\begin{equation}
    E'(s)
    =
    -\frac{1}{2}
    \frac{N'(s)}{N(s)}
    I(s).
\end{equation}
Since
\begin{equation}
    \frac{d}{ds}\log N(s)
    =
    \frac{N'(s)}{N(s)},
\end{equation}
we obtain
\begin{equation}
    \frac{dE}{d\log N}
    =
    -\frac{I(s)}{2}.
    \label{eq:hierarchical_slope}
\end{equation}
A further derivative gives
\begin{equation}
    \frac{d^2E}{d(\log N)^2}
    =
    -\frac{1}{2}
    \frac{I'(s)}
    {N'(s)/N(s)}.
    \label{eq:hierarchical_curvature}
\end{equation}
Because $N'(s)<0$, the condition $I'(s)<0$ is equivalent to
\begin{equation}
    \frac{d^2E}{d(\log N)^2}<0.
\end{equation}
Thus, continuous hierarchical localization occurs when $E$ is a concave function of
$\log N$.

The maximization of Eq.~\eqref{eq:hierarchical_F_energy_entropy} can be interpreted geometrically.
Writing
\begin{equation}
    y=\log N,
\end{equation}
we have
\begin{equation}
    \mathcal{F}(y)=E(y)+\frac{y}{\beta}.
\end{equation}
Changing $\beta$ therefore corresponds to changing the slope of a linear tilt added to the curve
$E(y)$. If $E(y)$ is concave, the maximizing point moves continuously as the tilt is varied.

If instead $E(y)$ contains a non-concave region, two separated hierarchical levels may become
degenerate before the intermediate levels are selected. Let the two competing levels be $s_1$
and $s_2$. Their coexistence condition reads
\begin{equation}
    E(s_1)+\frac{1}{\beta}\log N(s_1)
    =
    E(s_2)+\frac{1}{\beta}\log N(s_2),
\end{equation}
or equivalently
\begin{equation}
    \frac{1}{\beta}
    =
    -
    \frac{
        E(s_2)-E(s_1)
    }{
        \log N(s_2)-\log N(s_1)
    }.
    \label{eq:hierarchical_maxwell}
\end{equation}
If both extrema are locally stationary, Eq.~\eqref{eq:hierarchical_slope} additionally gives
\begin{equation}
    \left.
    \frac{dE}{d\log N}
    \right|_{s_1}
    =
    \left.
    \frac{dE}{d\log N}
    \right|_{s_2}
    =
    -\frac{1}{\beta}.
\end{equation}
Equations~\eqref{eq:hierarchical_maxwell} therefore correspond to a common-tangent construction
on the curve $E(\log N)$, analogous to a Maxwell construction. At the corresponding value of
$\beta$, the maximizing level jumps discontinuously from $s_1$ to $s_2$.

The continuum construction therefore separates two qualitatively distinct regimes. If
$I'(s)<0$ throughout the hierarchy, the selected level moves continuously toward smaller and
more correlated blocks as $\beta$ increases. If this monotonicity condition is violated, entire
ranges of hierarchical levels can be skipped, producing discontinuous jumps between different
localization scales.

\section{Continuous pattern distribution}
We can generalize the MHN (here we only consider the Euclidean case for simplicity) and use a distribution of patterns in the energy, namely:
\begin{equation}
    E(\boldsymbol{x}) = - \frac{1}{\beta} \log \int d^N \xi \, \rho(\boldsymbol{\xi}) e^{-\frac{\beta}{2} ||\boldsymbol{x} - \boldsymbol{\xi}||^2}
\end{equation}
with $\rho(\boldsymbol{\xi})$ being the patterns distribution. Writing $\rho(\boldsymbol{\xi})$ in a Boltzmann-like form $\rho(\boldsymbol{\xi}) \propto \exp(-V(\boldsymbol{\xi}))$ we get:
\begin{equation}
    E(\boldsymbol{x}) = \textrm{const.} - \frac{1}{\beta} \log \int d^N \xi \,  e^{ - V(\boldsymbol{\xi})-\frac{\beta}{2} ||\boldsymbol{x} - \boldsymbol{\xi}||^2}
\end{equation}
Changing variable $\boldsymbol{z} = \sqrt{\beta}(\boldsymbol{\xi} - \boldsymbol{x})$:
\begin{equation}
  E(\boldsymbol{x}) =   - \frac{1}{\beta} \log \left(\int d^N z e^{-\frac{||\boldsymbol{z}||^2}{2} - V(\boldsymbol{x} + \frac{1}{\sqrt{\beta}} \boldsymbol{z})}\right)
\end{equation}
Expanding around large $\beta$ (up to constants):
\begin{equation}
    E(\boldsymbol{x}) \approx \frac{1}{\beta} V(\boldsymbol{x}) + \frac{1}{2\beta^2} (\nabla^2 V(\boldsymbol{x}) - || \nabla V(\boldsymbol{x})||^2) + O(\beta^{-3})
\end{equation}
This shows that, at large $\beta$, the MHN energy coincides with the effective potential induced by the distribution, corrected by curvature-like terms. 
On the other hand, the small $\beta$ expansion yields:
\begin{equation}
    E(\boldsymbol{x}) \approx 
    \frac{1}{2} \|\boldsymbol{x}-\boldsymbol{\mu}_\xi\|^2
    - \frac{\beta}{2}
    \left[
        (\boldsymbol{x}-\boldsymbol{\mu}_\xi)^\intercal 
        \boldsymbol{\Sigma}_\xi
        (\boldsymbol{x}-\boldsymbol{\mu}_\xi)
        -
        \boldsymbol{m}_{3,\xi}^\intercal
        (\boldsymbol{x}-\boldsymbol{\mu}_\xi)
    \right],
\end{equation}
with $\boldsymbol{\mu}_\xi$ and $\boldsymbol{\Sigma}$ are the patterns mean and
\begin{equation}
    \boldsymbol{m}_{3,\xi}
    =
    \left\langle
        (\boldsymbol{\xi}-\boldsymbol{\mu}_\xi)
        \|\boldsymbol{\xi}-\boldsymbol{\mu}_\xi\|^2
    \right\rangle ,
\end{equation}
a contracted third central moment vector. This latter has an important role, as it probes -- already at first order in $\beta$ -- asymmetries of the distribution. The minimum in this case is approximately:
\begin{equation}
    \boldsymbol{x} - \boldsymbol{\mu}_\xi \approx (\boldsymbol{I} - \beta \boldsymbol{\Sigma}_\xi )^{-1} \frac{\beta}{2} \boldsymbol{m}_{3,\xi}
\end{equation}
which becomes unstable at $\beta_c \approx \lambda_{\rm max}(\boldsymbol{\Sigma}_\xi)^{-1}$, recalling the analogous scaling encountered for the finite dataset dynamics. 

We can make some notable explicit examples:
\begin{enumerate}
    \item Gaussian distribution. Here we simply have a smoothing of the quadratic potential, and no instability point occurs.
    \begin{equation}
        E(\boldsymbol{x}) = \frac{1}{2} (\boldsymbol{x}  - \boldsymbol{\mu})^\intercal \left(\boldsymbol{I} + \beta \boldsymbol{\Sigma}\right)^{-1} (\boldsymbol{x} - \boldsymbol{\mu}) + \textrm{const.}
    \end{equation}
    \item Gaussian Mixture Model. In this case, the energy reads:
    \begin{equation}
        E(\boldsymbol{x}) = - \frac{1}{\beta} \log \left( \sum_{r=1}^R \pi_r e^{-\frac{\beta}{2} (\boldsymbol{x}  - \boldsymbol{\mu}_r)^\intercal \left(\boldsymbol{I} + \beta \boldsymbol{\Sigma}_r\right)^{-1} (\boldsymbol{x} - \boldsymbol{\mu}_r) - \frac{1}{2} \log \det (\boldsymbol{I} + \beta \boldsymbol{\Sigma}_r)} \right)
    \end{equation}
We can immediately notice how each covariance matrix biases the entropic weight of its own component, and this bias disappears only for equal covariances. Indeed, in this latter case, apart from the different effective metric, we have a finite dataset MHN where we formally replaced the patterns by the component centers.
\end{enumerate}

\section{Flory-Huggins theory}
Our dual formulation potential for the MHN coincides (up to the overall sign) with the Gibbs free energy density for a multi-component mixture in the Flory-Huggins (FH) framework:
\begin{equation}
   f(\boldsymbol{\phi}) = k_B T \sum_{i=1}^C \phi_i \ln \phi_i + \frac{1}{2} \sum_{i,j=1}^C \chi_{ij} \phi_i \phi_j
\end{equation}
where the solute interaction matrix replaces the Gram matrix of the MHN. Of particular interest is the spatially extended version of the free energy:
\begin{equation}
    F = \int d^d x \left[ f(\boldsymbol{\phi}(\boldsymbol{x})) + \sum_{i=1}^C \frac{\kappa_i}{2} (\nabla \phi_i)^2  \right]
\end{equation}
This form of the FH theory endowed with an uncorrelated quenched disorder in the $\chi_{ij}$ has been studied in \cite{bartolucci2026metastabilityripeningmulticomponentliquid}, mapping directly to the case of i.i.d. patterns of the MHN. Under this interpretation, one could use the results presented in this paper to analyze multi-component mixtures or, at the same time, leverage the spatially extended version of the FH as a pattern retrieval model for, say, images. In this latter case, patterns, instead of being one particular image, could represent visual features that, when spatially combined, result in an image. We leave this exploration for future work.
\newpage
\section{Supplementary Figures}
\begin{figure*}[h]
    \centering
\includegraphics[width=\linewidth]{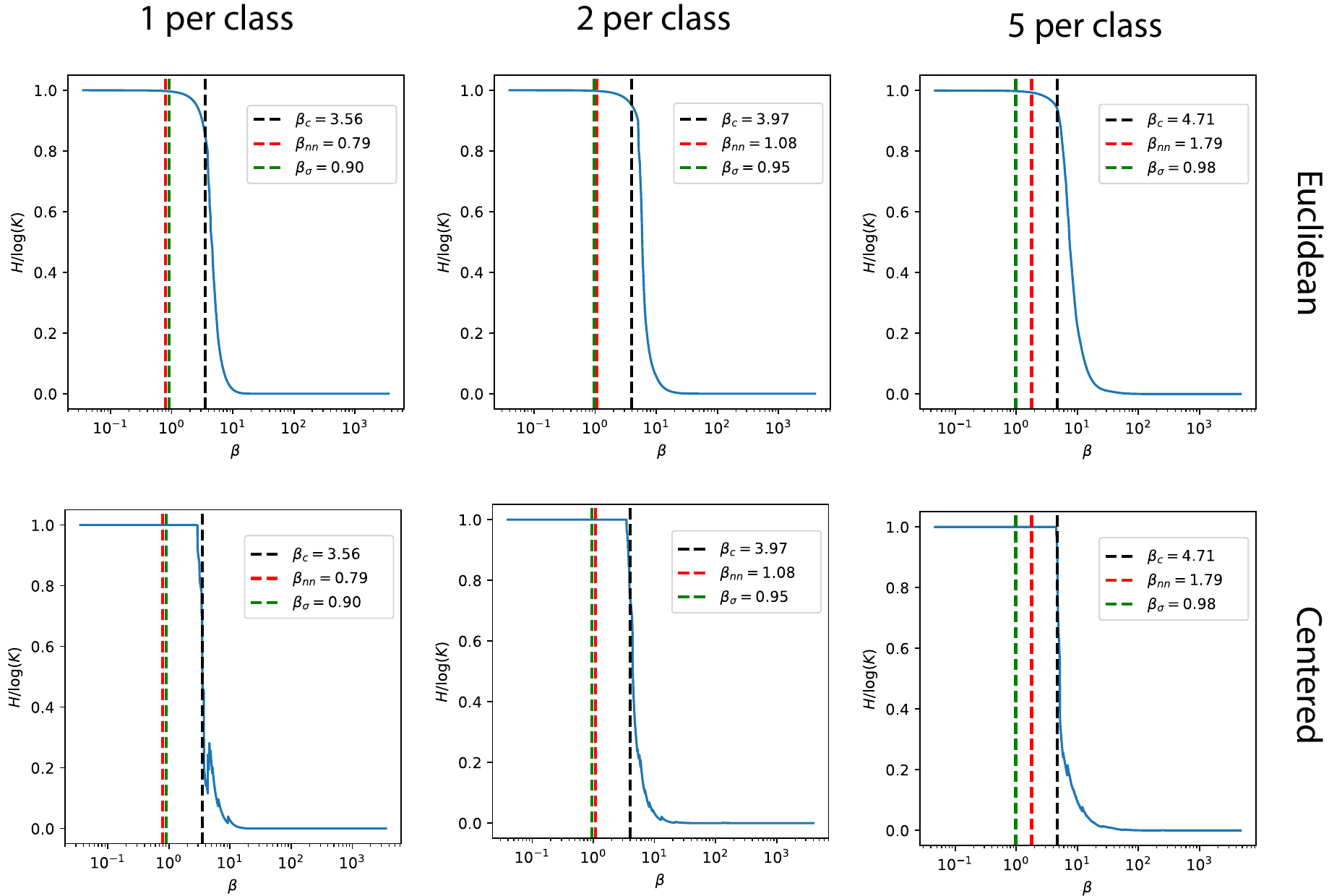}
    \caption{Normalized entropies of the quenched fixed points for the MNIST dataset. We report the Euclidean and centered cases. In the centered case, especially at low pattern number, a fixed point previously inaccessible can appear, manifesting in a non-monotonic behavior of the entropy as a function of $\beta$. The more patterns are introduced, the more the embedding space becomes crowded with patterns, thus smoothing the entropy.  }
    \label{fig:mnist_entropies}
\end{figure*}

\putbib[references]

\end{bibunit}

\end{document}